# High Thermoelectric Cooling Performance of Junction Thermoelectric Transistors

Chen Tang[‡1], Bohang Nan[‡1], Xiaodong Liu[2], Guiying Xu[1]*

1 Beijing Municipal Key Lab of Advanced Energy Materials and Technology, School of
  Materials Science and Engineering, University of Science and Technology Beijing,
  Beijing 100083, China.

2 Comprehensive Cancer Center of University of Puerto Rico, San Juan, Puerto Rico, 00921, USA

‡ Equal contribution.

* Corresponding Author: xugy@mater.ustb.edu.cn

## Abstract

To achieve high performance thermoelectric materials and devices, thermoelectric transistors, which integrate thermoelectric effects with transistor technology, represent a promising approach. Here p type $Bi_{0.5}Sb_{1.5}Te_3$ and n type $Bi_2Te_{2.97}Se_{0.03}$ are used as the constituent materials for an NPN transistor. By applying forward bias to the emitter and reverse bias to the collector to form a common-base triode configuration, the thermoelectric effect, transistor effect, and interfacial effects within the NPN heterostructure are coupled. This NPN heterostructure induces temperature increase (heat release) at the forward biased end and temperature decrease (heat absorption) at the reverse biased end. Therefore, this device becomes a new type of thermoelectric transistor cooler. Furthermore, a DC equivalent circuit method is introduced to analyze the cooling performance of the thermoelectric transistor cooler. The results show that when the emitter voltage of the NPN thermoelectric transistor cooler is 0.01925V, the collector voltage is 0.2124V, the corresponding base region width is 0.17nm, the maximum temperature difference of 242.89K can be obtained. Even if the base region width is limited to a level of 10 nanometers, for example when the base region width is 12.78nm, the maximum temperature differenced can still reach 174.15K. The research results above indicate that the thermoelectric transistor, which synergistically combines the thermoelectric effect with transistor technology, can effectively enhance the maximum temperature difference that typical thermoelectric cooling can achieve.

**Keywords:** Thermoelectric Transistor, NPN Heterostructure, Thermoelectric Cooling

## 1 Introduction

The continuous demand for energy-efficient cooling solutions, combined with environmental concerns, highlights the importance of exploring innovative methods to reduce global energy consumption and greenhouse gas emissions. Therefore, thermoelectric (TE) cooling has become one of the preferred solutions that attracts much attention.

Thermoelectric cooling, also called semiconductor refrigeration or thermoelectric refrigeration, is a new refrigeration technology developed in the late 1950s. It achieves refrigeration through the Peltier effect by directly using electrical energy. The Peltier effect was discovered by J. Peltier [1] in 1834, referring to the phenomenon that when electric current flows through the junction of two different materials, heat is absorbed from or released into the external environment. In practical applications, thermoelectric cooling possesses the advantages of: fast response, no pollution, no noise, no moving parts, small size, light weight, strong adaptability (adjustable in size, shape, temperature, power, etc.), high reliability (no wear, basically no maintenance required), high mechanical strength, strong impact resistance, long service life, coefficient of performance not depending on capacity, easy speed control, reversible operation [2,3]. As a result, thermoelectric cooling has been widely applied.

Despite having significant advantages in specific application domains such as aerospace, electronic devices, and healthcare, thermoelectric cooling technology is still constrained for large-scale commercialization by its relatively low conversion efficiency [4]. In recent years, scholars have pursued three primary strategies to enhance the performance of thermoelectric coolers:

First, enhancing the thermoelectric properties of materials [5,6]. The performance of thermoelectric materials is principally determined by the dimensionless figure of merit $ZT = S^2\sigma T/\kappa$, where $S$, $\sigma$, $T$, and $\kappa$ denote the Seebeck coefficient, electrical conductivity, absolute temperature, and total thermal conductivity (comprising lattice contribution $\kappa_l$ and electronic contribution $\kappa_e$) respectively [7-12]. Material optimization approaches include alloying [13-15], synthesis of nanocomposite structures [16-19], low-dimensional engineering [20-24]. Second, geometric and structural optimization of thermoelectric devices. Currently, the mainstream thermoelectric cooling device structures are three types: unipolar/layered devices constructed from series/parallel-connected π-type thermoelements (fundamental N/P-type material units), multistage thermoelectric coolers, Y-shaped structural devices [25]. Third, exploring novel concepts and operational principles – the primary focus of this research.

The combination of thermoelectric principles and transistor technology has attracted widespread attention, including electric-field-driven thermoelectric devices [26-30] and purely temperature-gradient-driven junction thermoelectric transistors [31-33]. Research indicates that in electric-field-driven thermoelectric devices, under the combined action of temperature gradient and gate voltage, thermoelectric performance can be optimized by independently regulating internal material parameters and decoupling the Seebeck coefficient and electrical conductivity [26]. Bejenari et al. used theoretical models to study the influence of gate voltage on the thermoelectric properties of $Bi_2Te_3$ nanowires [27]. The results indicated that the ZT of 7nm-thick $Bi_2Te_3$ nanowires could reach up to 3.4. Subsequently, experimental results from Qin et al. showed that under gate voltage, the ZT values of n-type and p-type $Bi_2Te_3$ thin films could achieve 1.22 and 1.02, respectively [28]. Through experimental measurements, Yang et al. discovered that the power factor ($S^2\sigma$) of 2D-$Bi_2O_2Se$ could exceed 400 $\mu Wm^{-1}K^{-2}$ due to gate voltage effects [29]. Additionally, Wu et al. found that under gate voltage, 2D $Nb_2SiTe_4$-based transistors realized the Seebeck diode effect at room temperature [30].

Thermoelectric junction transistors based on PNP/NPN heterostructures demonstrate substantial potential in thermoelectric by overcoming the inherent limitations of strong parameter coupling in conventional systems. With advancing applications of P-N junctions in thermoelectricity, research confirms that interfacial effects at P-N junctions enable significantly enhanced Seebeck coefficients near the depletion layer [31-36]. In 1986, Balmush et al. postulated that voltages across P-N junctions under non-isothermal conditions could increase dramatically near depletion regions [37]. Subsequent theoretical work in the 1990s by Zakhidov, Ravich, and collaborators demonstrated through detailed calculations that P-N junctions could substantially elevate both Seebeck coefficient and ZT values via interfacial phenomena [38-40]. Nan et al. systematically investigated the thermoelectric conversion performance of $(Bi_{1-x}Sb_x)_2(Te_{1-y}Se_y)_3$-based PNP junction thermoelectric transistors [31-33]. Through optimizing the carrier concentrations of the emitter, base, and collector, as well as the device dimensions, and utilizing thermal equilibrium equations, the PNP thermoelectric junction transistor achieves a maximum output power of 17.8mW under a 50K temperature difference drive, with a corresponding conversion efficiency of 8.69%, under maximum power output mode without considering the Thomson effect [32]. This efficiency would require conventional thermoelectric generators to achieve ZT > 9, underscoring the substantial potential of junction thermoelectric transistors for breakthrough performance enhancement.

Accordingly, in this paper, P-type $Bi_{0.5}Sb_{1.5}Te_3$ and N-type $Bi_2Te_{2.97}Se_{0.03}$ are used to construct NPN junction thermoelectric transistors. Positive and negative bias voltages are applied to the emitter and collector respectively to regulate the thermoelectric effect, interface effect and transistor effect of the junction-type thermoelectric transistor, thereby decoupling material parameters and improving the thermoelectric cooling performance. A systematic study is conducted on the output performance of the thermoelectric transistor cooling device.

## 2. Principles and Analytical Methods
### 2.1 Working principle and material selection of TE transistor cooling devices

The working principle of the general thermoelectric cooling device or the Peltier effect in semiconductor materials can be explained by Figure 1[41]. As can be seen from Figure 1, when the current flows from the N type region to the P type region, the charge carriers continuously flow out at the pn junction (1). Therefore, new charge carriers need to be generated accordingly, which requires energy consumption, thus causing the temperature to decrease. When the charge carriers flow to the pn junction (2), the two types of charge carriers move towards each other, resulting in the recombination of electrons and holes, thereby releasing energy and causing the temperature to rise. It can also be seen from Figure 1 that the pn junction (1) is reverse-biased, and the pn junction (2) is forward-biased. That is, reverse bias causes the interface temperature to decrease, and forward bias causes the interface temperature to increase. Moreover, the currents passing through the two pn junctions in Figure 1 cannot be independently controlled because they share a single power supply.

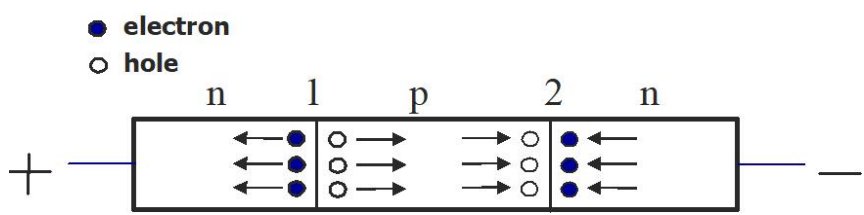

**Figure 1** Schematic diagram of the Peltier effect of semiconductor materials

Here, the rectangular structure design of the NPN heterojunction is shown in Figure 2 (a). Along the X direction, $W_{n1}$, $W_p$, and $W_{n2}$ represent the widths of $N_1$-type, P-type, and $N_2$-type materials, respectively. In the YZ plane, the cross-sectional area of the NPN heterojunction is A. The P-type region is fully depleted to eliminate the effects of bipolar diffusion. Therefore, there are no free minority carriers (holes) in this P-type region. And the dimensions of $W_{n1}$ and $W_{n2}$ are at the nanoscale. The circuit connection structure diagram of this NPN junction transistor used for refrigeration is shown in Figure 2 (b). The voltages applied between the $N_1$-P junction and the $N_2$-P junction ensure that the $N_1$-P junction is in reverse bias while the $N_2$-P junction is in forward bias. When current flows through the circuit, the temperature at the right end of the $N_2$ region will increase, while the temperature at the left end of the $N_1$ region will decrease, thereby creating a temperature difference between the two ends of the the $N_1PN_2$ junction thermoelectric transistor. Compared with Figure 1, the fundamental difference between the two figures lies in that the voltages of the two pn junctions in Figure 2 can be independently controlled. Therefore, by adjusting their voltages, the temperature of the $N_2$-P junction at the high-temperature end in Figure 2 can be lowered, which is more conducive to improving the cooling effect of the semiconductor cooling device.

The working principle is as follows: By controlling the carrier concentration of emitter $N_2$ and base P, as well as $V_2$, the current through the interface between collector $N_1$ and base P can be regulated. By adjusting the carrier concentration of $N_1$ and $V_1$, the cooling capacity can be controlled. This is because in a junction transistor, the emitter current $J_e$ is approximately equal to the collector current $J_c$, i.e., $J_e \approx J_c$. Therefore, by optimizing and controlling the carrier concentration of emitter $N_2$, base P, and collector $N_1$, as well as the applied voltages $V_2$ and $V_1$, the cooling capacity of the collector can be maximized.

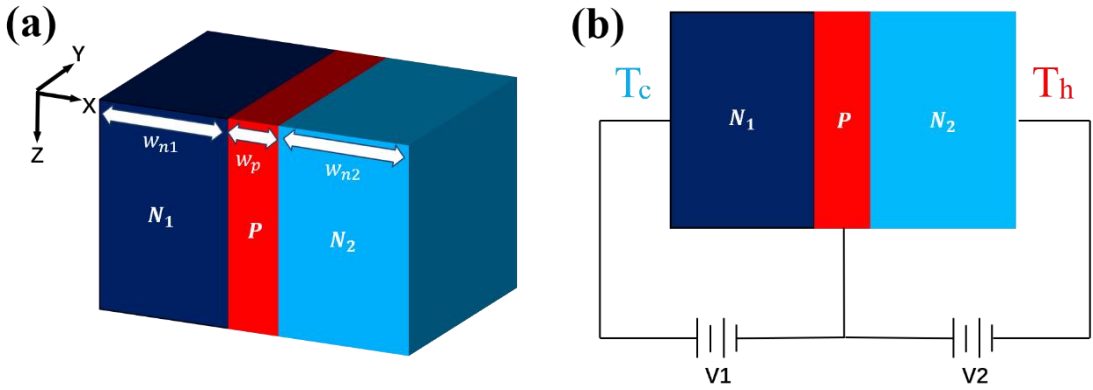

**Figure 2** (a) Schematic diagram of the thermoelectric transistor structure; (b) Circuit configuration of the thermoelectric transistor

In this work, $N_1$-$Bi_2Te_{2.97}Se_{0.03}$, P-$Bi_{0.5}Sb_{1.5}Te_3$, and $N_2$-$Bi_2Te_{2.97}Se_{0.03}$ are employed as the collector, base, and emitter materials, respectively, with their essential parameters summarized in Table 1.

Table 1 Material Parameters of TE transistors

| Parameters | Emitter | Base | Collector | References |
|---|---|---|---|---|
| $n_i(\times 10^{25} m^{-3})$ | 0.16 | 0.14 | 0.16 | [32] |
| k(W/mK) | 1.7474 | 1.7474 | 1.7474 | [42,43] |
| $\varepsilon_r$ | 90 | 90 | 90 | [40] |
| $\mu(m^2/Vs)$ | 0.12 | 0.0275 | 0.12 | [44] |
| L(μm) | 40 | 30 | 40 | [45] |
| $m^*/m_e$ | 1.39 | 1.44 | 1.39 | [46,47] |

## 2.2 Carrier concentration analysis

In thermoelectric transistors, since the P type region is completely depleted, only the charge carriers (electrons) in the N-type region can move through the NP or PN junction. In this work, N-type $Bi_2Te_{2.97}Se_{0.03}$ is used as the N type thermoelectric material; therefore, the majority electrons contribute more to the thermoelectric process than the minority holes [48]. Hereafter, we focus on the electron concentration distribution of N-type $Bi_2Te_{2.97}Se_{0.03}$. The distribution of electrons can be obtained through a theoretical analysis method. This method consists of three equations: the current density expression for DC current transport [49], the one-dimensional Poisson equation [47], and the current continuity equation [50]:

The current density equation governing DC charge transport can be expressed as formula (1):

$$J_n(x) = \sigma_n(x)[E(x) - S_n(x)\nabla T] \qquad (1)$$

The one-dimensional Poisson equation can be expressed as formula (2):

$$E(x) = \frac{d\varphi(x)}{dx} \qquad (2a)$$

$$\frac{d^2\varphi(x)}{dx^2} = -\frac{e}{\varepsilon_r \varepsilon_0}(n(x) - Nd) \qquad (2b)$$

The current continuity equation can be expressed as formula (3):

$$\frac{\partial n}{\partial t} = -\frac{1}{e}\frac{\partial J_n(x)}{\partial x} + (G_n - R_n) \qquad (3)$$

Where $J_n(x)$ represents the current density, $\sigma_n(x)$ and $S_n(x)$ denote the electrical conductivity and Seebeck coefficient of N-type $Bi_2Te_{2.97}Se_{0.03}$, $\nabla T$ signifies the temperature gradient, $E(x)$ stands for the electric field, $\varphi(x)$ indicates the electric potential, e is the charge, $\varepsilon_r$ is the relative dielectric constant, $\varepsilon_0$ is the vacuum dielectric constant, $n(x)$ is the electron concentration distribution in N-type $Bi_2Te_{2.97}Se_{0.03}$, $N_d$ represents the acceptor doping concentration of N-type $Bi_2Te_{2.97}Se_{0.03}$, $G_n$ represents the generation rate of electron carriers, and Rn denotes the recombination rate of electron carriers.

Since the N-type $Bi_2Te_{2.97}Se_{0.03}$ used in this work can be approximated as a degenerate semiconductor, the conductivity $\sigma_n(x)$ and the seebeck coefficient $S_n(x)$ can be expressed as formula (4) [47][51]:

$$\sigma_n(x) = n(x)e\mu_n \qquad (4a)$$

$$S_n(x) = \left[\frac{8\pi^{\frac{8}{3}} m_n^* k_B^2 \left(r + \frac{3}{2}\right)}{3^{\frac{5}{3}} h^2 e}\right] \frac{T(x)}{n(x)^{\frac{2}{3}}} \qquad (4b)$$

Where $\mu_n$ is the electron mobility, $k_B$ is the Boltzmann constant, e is the electron charge, r = -1/2 is the scattering factor, $m_n^*$ is the effective mass of the carrier, h is the Planck constant, and $T(x)$ is the temperature distribution.

At the high-temperature end (the right end of $N_2$), a constant temperature boundary condition is adopted. Assuming that the temperature in the N-type $Bi_2Te_{2.97}Se_{0.03}$ is linearly distributed, with the temperature at the high-temperature end (the right end of $N_2$) being $T_h$ and the temperature at the low-temperature end (the left end of $N_1$) being Tc, the temperature distribution can be expressed as:

$$T(x) = -\frac{T_h - T_c}{l_x}x + T_h \qquad (5)$$

Since the simulation temperature is lower than the intrinsic excitation temperature of the N-type $Bi_2Te_{2.97}Se_{0.03}$ and it can be assumed that the electron generation rate equals the recombination rate, thus the electron concentration remains unchanged.

$$\int_0^{l_x} N_d\, dx = \int_0^{l_x} n(x)dx \tag{6}$$

Then, formula (1b) becomes:

$$E(x) = \frac{d\varphi(x)}{dx} = \int_0^{l_x} -\frac{e}{\varepsilon_r \varepsilon_0}(n(x) - N_d)dx = 0 \tag{7}$$

Substituting equation (7) into equation (1a) yields

$$J_n(x) = -\sigma_n(x)S_n(x)\nabla T \tag{8}$$

Meanwhile, assuming that the temperature of the N-type $Bi_2Te_{2.97}Se_{0.03}$ (including the $N_1$ and $N_2$ regions) is in a dynamically stable distribution state, then

$$\frac{\partial n}{\partial t} = -\frac{1}{q}\frac{\partial J_n(x)}{\partial x} = 0 \tag{9}$$

Finally, based on equations (4a), (4b), (5), (8), and (9), the electron concentration distribution in N-type $Bi_2Te_{2.97}Se_{0.03}$ can be calculated.

In the thermoelectric transistor of this work, both the $N_1$ and $N_2$ regions are composed of N-type $Bi_2Te_{2.97}Se_{0.03}$. Since the P-type region is extremely thin (only a few nanometers in thickness), it is assumed that its temperature remains constant at $T_0$. The temperature distributions are as follows:

$$T_1(x_1) = -\frac{T_0 - T_c}{w_{n1}}x_1 + T_0 \tag{10}$$

$$T_2(x_2) = -\frac{T_h - T_0}{w_{n2}}x_2 + T_h \tag{11}$$

Where $w_{n1}$ and $w_{n2}$ represent the widths of the $N_1$ and $N_2$ regions, respectively.

Based on equations (4a), (4b), (5), (8), (9), (10), and (11), the electron concentration distributions in the $N_1$ and $N_2$ regions are given by:

$$N_1(x_1) = 2N_{d1}w_{n1}/[\left(\frac{T_0 w_{n1}}{T_c - T_0}\right)^{-2} - \left(\frac{T_c w_{n1}}{T_c - T_0}\right)^{-2}] * \left(x_1 + \frac{T_0 w_{n2}}{T_c - T_0}\right)^{-3} \tag{12}$$

$$N_2(x_1) = 2N_{d2}w_{n2}/[\left(\frac{T_h w_{n2}}{T_0 - T_h}\right)^{-2} - \left(\frac{T_0 w_{n2}}{T_0 - T_h}\right)^{-2}] * \left(x_2 + \frac{T_h w_{n2}}{T_0 - T_h}\right)^{-3} \tag{13}$$

Thus, the electron concentration n1 in the $N_1$ region at the $N_1$-P interface and the electron concentration n2 in the $N_2$ region at the $N_2$-P interface are respectively:

$$n_1 = N_1(0) \tag{14}$$

$$n_2 = N_2(w_{n2}) \tag{15}$$

### 2.3 Operating States of the TE Transistor

In this thermoelectric transistor, the $N_1$, P and $N_2$ regions are used as the collector, base, and emitter, respectively. The applied voltages $V_1$ and $V_2$ divide the thermoelectric transistor into reverse-biased and forward-biased sections.

In the reverse-biased section of the thermoelectric transistor, the P-type region is fully depleted. Consequently, the Seebeck voltage $V_{s1}$ and Seebeck coefficient $S_{n1}$ in the $N_1$ region are given by:

$$V_{s1} = S_{n1}(T_0 - T_c) \tag{16}$$

$$S_{n1} = \left[\frac{8\pi^{\frac{8}{3}} m_n^* k_B^2 \left(r + \frac{3}{2}\right)}{3^{\frac{5}{3}} h^2 e}\right] \frac{\frac{(T_c + T_0)}{2}}{N_{d1}^{\frac{2}{3}}} \tag{17}$$

Where $k_B$ is the Boltzmann constant, e is the electron charge, $r = -1/2$ is the scattering factor, $m_n^*$ is the carrier effective mass, and h is the Planck constant.

The built-in voltage $V_{d1}$ of the $N_1$-P junction is given by:

$$V_{d1} = \frac{k_B T_0}{e} \ln \frac{n_1 p_1}{n_{ni} n_{pi}} \tag{18}$$

Where $n_{ni}$ and $n_{pi}$ represent the intrinsic carrier concentrations of N-type $Bi_2Te_{2.97}Se_{0.03}$ and P-type $Bi_{0.5}Sb_{1.5}Te_3$, respectively. $n_1$ denotes the electron concentration in the $N_1$ region at temperature $T_0$, $p_1$ is the hole concentration in the P-type region. Since the P-type region is at the nanoscale, it is assumed that $p_1$ is equal to the doping concentration $P_a$ of the P-type region.

Similarly, the Seebeck voltage $V_{s2}$ and Seebeck coefficient $S_{n2}$, in the $N_2$ region, and the built-in voltage $V_{d2}$ at the $N_2$-P interface are given by:

$$V_{s2} = S_{n2}(T_h - T_0) \tag{19}$$

$$S_{n2} = \left[\frac{8\pi^{\frac{8}{3}} m_p^* k_B^2 \left(r + \frac{3}{2}\right)}{3^{\frac{5}{3}} h^2 e}\right] \frac{\frac{(T_h + T_0)}{2}}{N_{d2}^{\frac{2}{3}}} \tag{20}$$

$$V_{d2} = \frac{k_B T_0}{e} \ln \frac{n_2 p_2}{n_{ni} n_{pi}} \tag{21}$$

Where $S_{n2}$ represents the Seebeck coefficient in the $N_2$ region, $n_2$ denotes the electron concentration in the $N_2$ region at temperature $T_0$ (which equals its acceptor doping concentration), and $p_2$ is the hole concentration in the P-type region. Since the P-type region is at the nanoscale, it is assumed that $p_2$ is equal to the doping concentration $P_a$ of the P-type region.

## 2.4 Working Principle and DC Equivalent Circuit Diagram of the TE Transistor Cooling Devices

In the thermoelectric transistor circuit , the base-collector leakage current and Early voltage effects are neglected. The equivalent DC circuit is illustrated in Fig 2. The left section represents the reverse-biased part, where $V_{s1}$ is the Seebeck voltage of the $N_1$-P region, $R_1$ is the total resistance of the $N_1$-P region including the resistances of both $N_1$ and P regions as well as the interface resistance, and $V_{d1}$ is the interface voltage between $N_1$ and P. The right part corresponds to the forward-biased section, where $V_{s2}$ is the Seebeck voltage generated by the $N_2$-P region, $R_2$ is the total resistance of the $N_2$-P region including the resistances of both $N_2$ and P regions as well as the interface resistance, and $V_{d2}$ is the interface voltage between $N_2$ and P. Additionally, the contact resistance between the thermoelectric transistor and electrodes is neglected in this model.

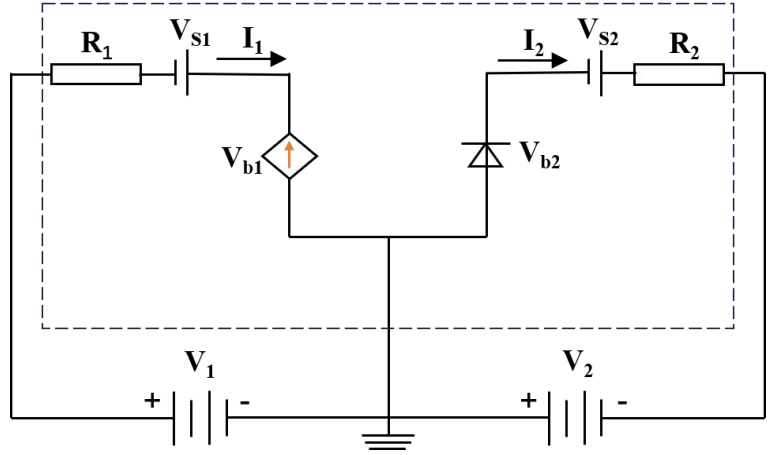

**Figure 3**. DC equivalent circuit diagram of the thermoelectric transistor cooling device.

The internal equivalent circuit of the TE transistor is represented within the black dashed box in the Figure 3。

According to Kirchhoff's laws, the relationships between these electrical parameters in the thermoelectric transistor can be expressed as:

$$V_1 = I_1 R_1 + V_{s1} + V_{d1} \tag{22}$$

$$V_2 = I_2 R_2 + V_{s2} - V_{d2} \tag{23}$$

$$I_1 = I_2 = J * A \tag{24}$$

$$R_1 = R_{n1} + R_{p1} + R_{j1} \tag{25}$$

$$R_2 = R_{n2} + R_{p2} + R_{j2} \tag{26}$$

Additionally

$$R_{n1} = \frac{1}{N_{d1} e \mu_n} * \frac{w_{n1}}{A} \tag{27}$$

$$R_{p1} = \frac{1}{P_a e \mu_p} * \frac{w_{p1}}{A} \tag{28}$$

$$R_{n2} = \frac{1}{N_{d2} e \mu_n} * \frac{w_{n2}}{A} \tag{29}$$

$$R_{p2} = \frac{1}{P_a e \mu_p} * \frac{w_{p2}}{A} \tag{30}$$

$$R_{j1} = \frac{k_B T_0}{A e^2} \left( \frac{D_{p1} n_{ni}}{L_p} + \frac{D_{n1} n_{pi}}{L_n} \right)^{-1} \tag{31}$$

$$R_{j2} = \frac{k_B T_0}{A e^2} \left( \frac{D_{p2} n_{ni}}{L_p} + \frac{D_{n2} n_{pi}}{L_n} \right)^{-1} \tag{32}$$

Where J represents the current density, A denotes the cross-sectional area of the thermoelectric transistor, $L_p$ and $L_n$ are the diffusion lengths of P-type $Bi_{0.5}Sb_{1.5}Te_3$ and N-type $Bi_2Te_{2.97}Se_{0.03}$, respectively, and $\mu_p$ and $\mu_n$ are the charge carrier mobilities of P-type $Bi_{0.5}Sb_{1.5}Te_3$ and N-type $Bi_2Te_{2.97}Se_{0.03}$. The diffusion coefficients $D_{p1}$, $D_{n1}$, $D_{p2}$, and $D_{n2}$ correspond to the reverse-biased and forward-biased regions of P-type $Bi_{0.5}Sb_{1.5}Te_3$ and N-type $Bi_2Te_{2.97}Se_{0.03}$,

respectively. These diffusion coefficients (D) can be derived from the Einstein relation:

$$D = \frac{k_B T}{e} \mu \tag{33}$$

Since the thermoelectric cooling process belongs to high current injection, the curren density formula is given by:

$$J = 2e\left(\frac{D_p n_{ni}}{L_p} + \frac{D_n n_{pi}}{L_n}\right) \exp\left(\frac{eV_2}{2k_B T_0}\right) \tag{34}$$

Since P$_1$ and P$_2$ are completely depleted, w$_{p1}$ and w$_{p2}$ can be expressed as

$$w_{p1} = \sqrt{\frac{2\varepsilon_1 \varepsilon_0 V_{d1}}{e} * \frac{n_1}{P_a * (P_a + n_1)}} \tag{35}$$

$$w_{p2} = \sqrt{\frac{2\varepsilon_2 \varepsilon_0 V_{d2}}{e} * \frac{n_2}{P_a * (P_a + n_2)}} \tag{36}$$

$$W_p = W_{p1} + W_{p2} \tag{37}$$

Where ε$_1$ and ε$_2$ represent the relative permittivity of the P$_1$ and P$_2$ regions, respectively, while ε$_0$ denotes the vacuum permittivity.

## 2.5 Thermoelectric Transistor Cooling Performance
### 2.5.1 Thermoelectric Transistor Conversion Efficiency

The conversion efficiency (η) of a thermoelectric cooler is defined as:

$$\eta = \frac{Q_c}{P} \tag{38}$$

Where η is the conversion efficiency, Q$_c$ denotes the heat absorption at the cold end, and P stands for the input electrical power.

When considering the coupling between the Seebeck effect and interface effects in NPN heterostructures, Q$_c$ can be expressed as:

$$Q_c = (S_{n1} T_0 + V_{d1}) I_1 - \frac{1}{2} I_1^2 R_1 - \frac{1}{2} I_2^2 R_2 - k(T_h - T_c) \tag{39}$$

$$k = \frac{A_N}{l_N} \lambda_N + \frac{A_P}{l_P} \lambda_P \tag{40}$$

In the equation above, λ$_P$ and λ$_N$ represent the thermal conductivities of the P-type and N-type materials, respectively, while A$_P$, l$_P$ and A$_N$, l$_N$ are the cross-sectional areas and lengths of the P-type and N-type materials, respectively.

### 2.5.2 Cooling Temperature Difference

The input power P of the refrigeration device can be expressed as:

$$P = I_1 V_1 + I_2 V_2 \tag{41}$$

By substituting equations (37) and (39) into the equation of efficiency (38), we can obtain

$$\eta = \frac{(S_{n1} T_0 + V_{d1}) I_1 - \frac{1}{2} I_1^2 R_1 - \frac{1}{2} I_2^2 R_2 - k(T_h - T_c)}{I_1 V_1 + I_2 V_2} \tag{42}$$

Another important performance indicator of a thermoelectric cooler is the temperature difference that can be established across its two ends, that is

$$\Delta T = T_h - T_c \tag{43}$$

Using the thermal equilibrium equation of the cold end of the device (Equation (40)), the result can be obtained

$$\Delta T = \frac{(S_{n1}T_0 + V_{d1})I_1 - \frac{1}{2}I_1^2 R_1 - \frac{1}{2}I_2^2 R_2 - Q_c}{k} \tag{44}$$

When the cooler works with no external heat load ($Q_c$=0), the maximum temperature difference $\Delta T_{max}$ that the cooler can achieve can be obtained.

### 2.5.3 Calculation of $T_0$

Since the thermoelectric transistor dissipates heat only at the forward bias and absorbs heat at the reverse bias, the heat absorbed at the forward bias portion, $Q_{c2}$, equals the heat dissipated at the reverse bias portion, $Q_{h1}$. This can be expressed as

$$Q_{c2} = (\alpha_{n2}T_0 - V_{d2})I_2 - \frac{1}{2}I_2^2 R_2 - k(T_h - T_0) \tag{45}$$

$$Q_{c1} = (\alpha_{n1}T_0 + V_{d1})I_1 - \frac{1}{2}I_1^2 R_1 - k(T_0 - T_c) \tag{46}$$

$$P_1 = I_1 V_1 = I_1^2 R_1 + \alpha_{n1}(T_0 - T_c)I_1 + I_1 V_{d1} \tag{47}$$

$$Q_{h1} = Q_{c1} + P_1 = [\alpha_{n1}(2T_0 - T_c) + 2V_{d1}]I_1 + \frac{1}{2}I_1^2 R_1 - k(T_0 - T_c) \tag{48}$$

Based on equations (46) and (49), we can determine $T_0$, which is

$$T_0 = \frac{2V_{d1}I_1 + V_{d2}I_2 - \alpha_{n1}T_c I_1 + \frac{1}{2}(I_1^2 R_1 + I_2^2 R_2) + k(T_h + T_c)}{\alpha_{P2}I_2 - 2\alpha_{P1}I_1 + 2k} \approx \frac{T_h + T_c}{2} \tag{49}$$

During the calculation process, $T_0$ can be approximated as $(T_h + T_c)/2$.

### 2.6 Calculation Methods and Process

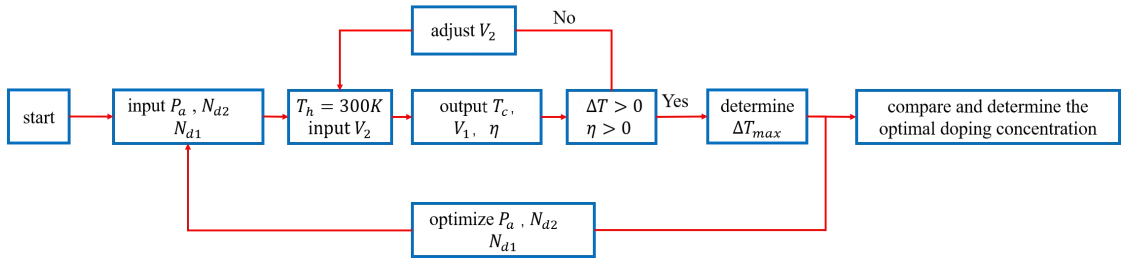

**Figure 4** Calculation flowchart of optimal doping concentration

The results of this work are computed using MATLAB R2020a. During the computation, $N_{d1}$ is set equal to $N_{d2}$, and $T_h$ was fixed at 300K. Figure 4 illustrates the schematic of the optimization design and computational process. Using the trial-and-error method, first determine the values of $P_a$, $N_{d1}$, and $N_{d2}$. Then, substitute the values of $P_a$, $N_{d1}$, $N_{d2}$, and $V_2$ into formula (43). Under the condition of $T_c < T_h$ and $\eta > 0$, adjust the value of $V_2$ and calculate the corresponding values of $V_1$, $\Delta T$, and $\eta$. Finally, obtain the allowable range of changes in $V_2$, $V_1$, $\Delta T$, and $\eta$. Based on the results above, the parameters $P_a$, $N_{d1}$, and $N_{d2}$ are optimized. The maximum temperature difference $\Delta T_{max}$ under different doping concentrations is calculated, respectively. After comparison, the optimal doping concentration was obtained.

## 3 Results and Discussion

In this work, the thickness of the thermoelectric transistor along the Y and Z directions is 0.1 mm, and the width of the $N_1$ and $N_2$ regions along the X direction is 0.06 mm. The high-temperature end temperature $T_h$ is maintained at 300 K, and it is assumed that the carrier concentrations of the emitter N1 and the collector N2 are $N_{d1}$ and $N_{d2}$, respectively, with $N_{d1} = N_{d2}$. The performance parameters of other materials are shown in Table 1. In this calculation, since the maximum temperature difference is obtained when $N_{d1} = N_{d2} = 1.5$–$3.5\times10^{25}$ m$^{-3}$ at relatively low $P_a$, the numerical simulation results for $N_{d1} = N_{d2} = 0.6$–$3.5\times10^{25}$ m$^{-3}$ are provided.

### 3.1 The calculation result of $N_{d1}=N_{d2}=0.6\times10^{25}$m$^{-3}$

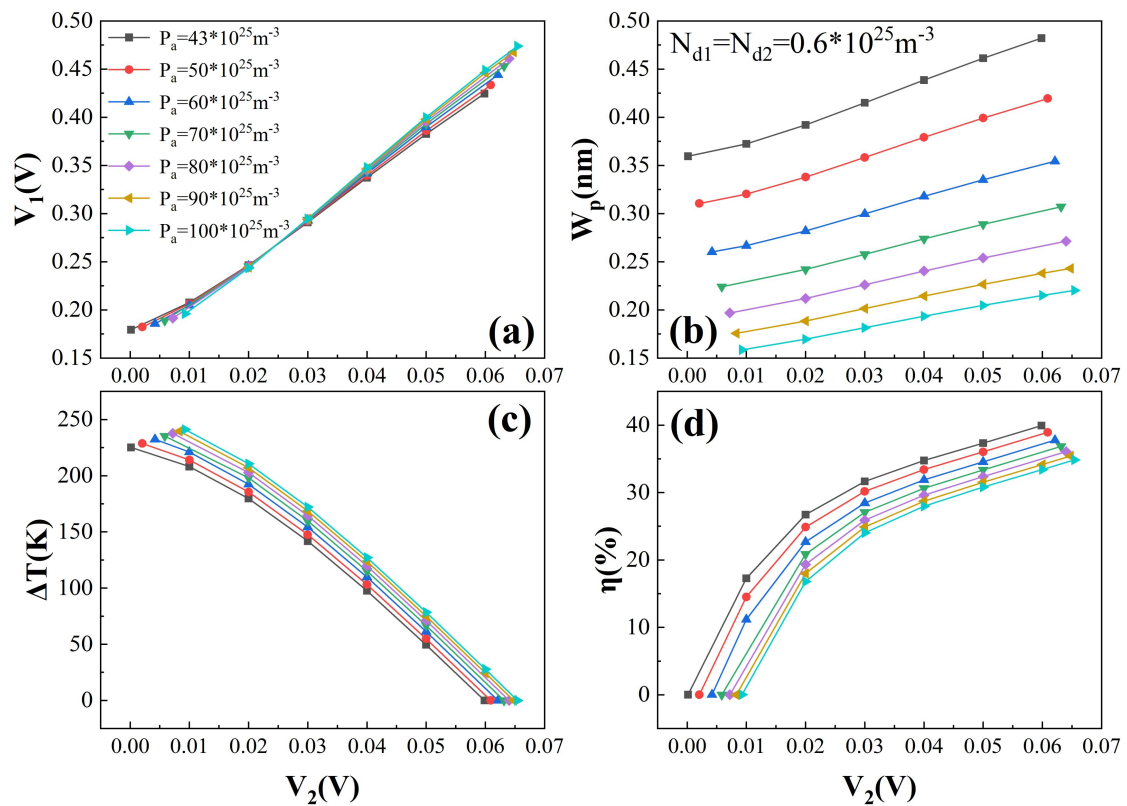

**Figure 5** $V_1$ (a), $W_p$ (b), $\Delta T$ (c), $\eta$ (d) as functions of $V_2$ and $P_a$ when $N_d = 0.6\times10^{25}$ m$^{-3}$

Figure 5 (a), (b), (c), and (d) show the variation laws of collector voltage $V_1$, base thickness $W_p$, temperature difference $\Delta T$ across the device, and thermoelectric conversion efficiency $\eta$ with respect to emitter voltage $V_2$ and base carrier concentration $P_a$ under N type region doping concentrations $N_{d1} = N_{d2} = 0.6\times10^{25}$ m$^{-3}$. From Figure 5 (a), under the same $P_a$, as the voltage $V_2$ applied to the emitter increases, the collector voltage $V_1$ increases. Additionally, as the base carrier concentration $P_a$ increases, the minimum and maximum allowable values of $V_2$ also increase, as do the initial and maximum reverse bias voltages $V_1$ that need to be applied to the collector. From Figure 5 (b), it can be observed that the base thickness $W_p$ decreases as the voltage $V_2$ applied to the emitter and the base carrier concentration $P_a$ increases. From Figure 5 (c), when Pa is constant, as $V_2$ increases, the temperature difference $\Delta T$ across the device decreases. When $V_2$ is constant, as $P_a$ increases, $\Delta T$ increases, and the maximum temperature difference $\Delta T_{max}$ across the device also increases with $P_a$. In Figure 5 (d), the thermoelectric conversion efficiency η decreases as $P_a$ increases when $V_2$ is constant and increases as $V_2$ increases when $P_a$ is constant. Combining Figures 5 (a), (b), (c), and (d), it can be seen that when $P_a$ is constant, the temperature difference $\Delta T$ across the device is maximized when $V_2$ is at its minimum allowable value, with corresponding minimum values of $V_1$ and $W_p$, and η = 0. Comparing the maximum temperature differences $\Delta T_{max}$ under different $P_a$ conditions, the maximum temperature difference across the device increases from $\Delta T_{max}$ = 225.17 K at $V_2$ = 0.00014 V, $V_1$ = 0.1794 V, and $P_a$ = 43×10$^{25}$ m$^{-3}$ to $\Delta T_{max}$ = 241.10 K at $V_2$ = 0.00928 V, $V_1$ = 0.1960 V, and $P_a$ = 100×10$^{25}$ m$^{-3}$. At these points, the base thickness $W_p$ is 0.36 nm and 0.16 nm, respectively.

## 3.2 The calculation result of $N_{d1}=N_{d2}=0.7\times10^{25}$m$^{-3}$

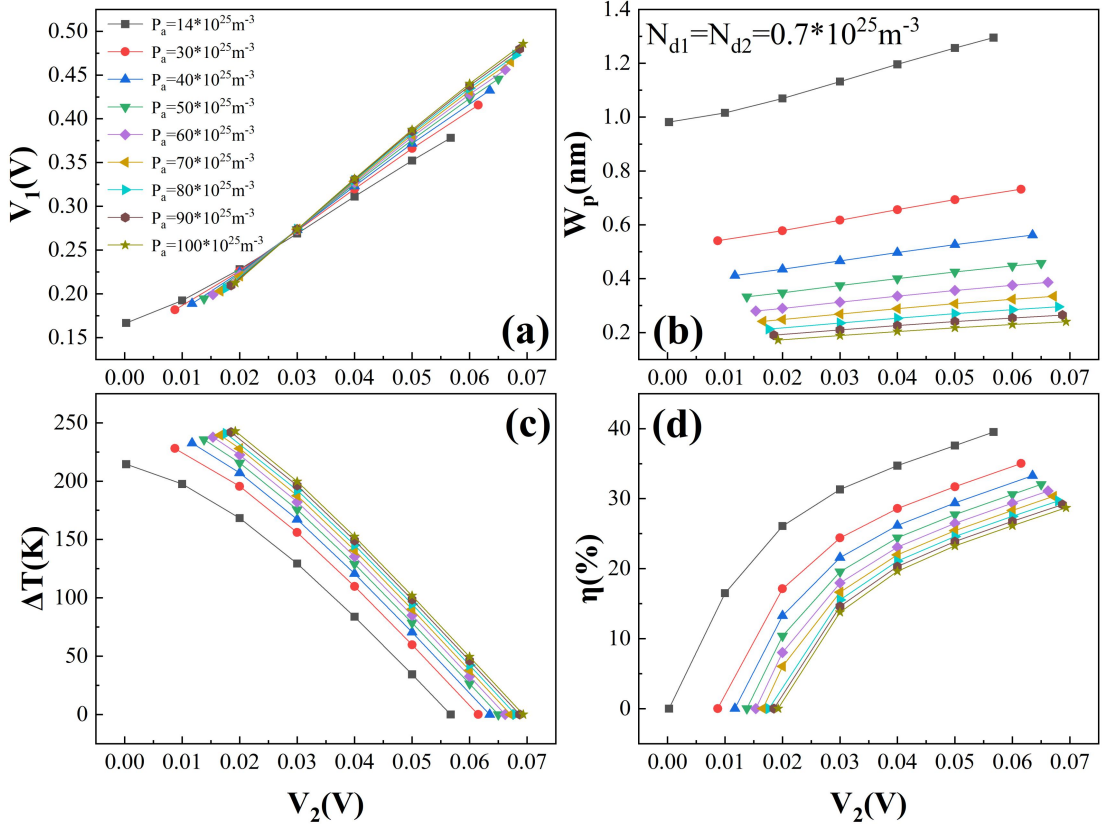

**Figure 6** $V_1$ (a), $W_p$ (b), $\Delta T$ (c), η (d) as functions of $V_2$ and $P_a$ when $N_d$ = 0.7×10$^{25}$ m$^{-3}$

Figure 6 (a), (b), (c), and (d) show the variation laws of collector voltage $V_1$, base thickness $W_p$, temperature difference $\Delta T$ across the device, and thermoelectric conversion efficiency η with respect to emitter voltage $V_2$ and base carrier concentration $P_a$ under N type region doping concentrations $N_{d1}$ = $N_{d2}$ = 0.7×10$^{25}$ m$^{-3}$. From Figure 6 (a), under the same Pa, as the emitter voltage $V_2$ increases, the collector voltage $V_1$ also increases. Additionally, as the base carrier concentration $P_a$ increases, the minimum and maximum allowable values of $V_2$ increase, as do the initial and maximum reverse bias voltages $V_1$ that need to be applied to the collector. From Figure 6 (b), it can be observed that the base thickness $W_p$ decreases as the emitter voltage $V_2$ and base carrier concentration $P_a$ increases. From Figure 6 (c), when $P_a$ is constant, the temperature difference $\Delta T$ across the device decreases as $V_2$ increases. When $V_2$ is constant, $\Delta T$ increases with $P_a$, and the maximum temperature difference $\Delta T_{max}$ across the device also increases with $P_a$. In Figure 6 (d), the thermoelectric conversion efficiency $\eta$ decreases as $P_a$ increases when $V_2$ is constant and increases as $V_2$ increases when $P_a$ is constant. Combining Figures 6 (a), (b), (c), and (d), it can be seen that when $P_a$ is constant, the temperature difference $\Delta T$ across the device is maximized when $V_2$ is at its minimum allowable value, with corresponding minimum values of $V_1$ and $W_p$, and $\eta = 0$. Comparing the maximum temperature differences $\Delta T_{max}$ under different $P_a$ conditions, the maximum temperature difference across the device increases from $\Delta T_{max} = 214.49$ K at $V_2 = 0.00028$ V, $V_1 = 0.1667$ V, and $P_a = 16 \times 10^{25}$ m$^{-3}$ to $\Delta T_{max} = 242.89$ K at $V_2 = 0.01925$ V, $V_1 = 0.2124$ V, and $P_a = 100 \times 10^{25}$ m$^{-3}$. At these points, the base thickness $W_p$ is 0.98 nm and 0.17 nm, respectively.

### 3.3 The calculation result of $N_{d1}=N_{d2}=0.8\times10^{25}$m$^{-3}$

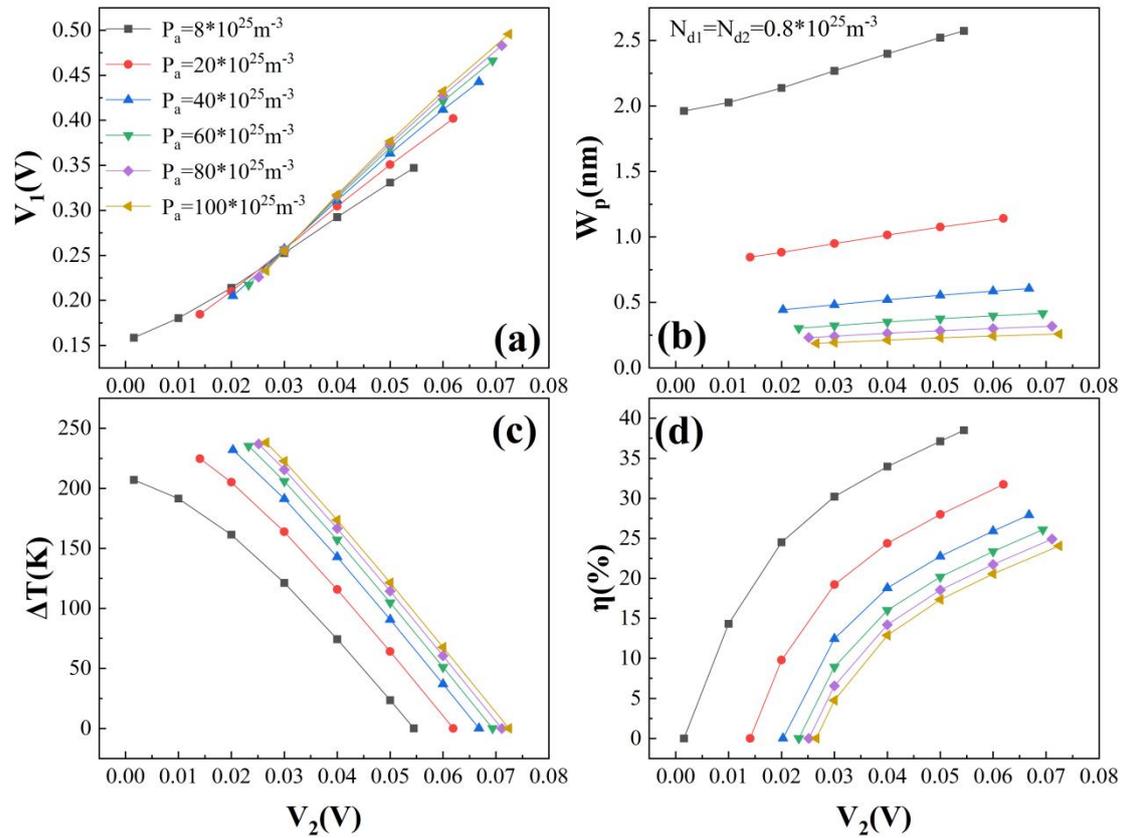

**Figure 7** $V_1$ (a), $W_p$ (b), $\Delta T$ (c), $\eta$ (d) as functions of $V_2$ and $P_a$ when $N_d = 0.8 \times 10^{25}$ m$^{-3}$

Figure 7 (a), (b), (c), and (d) show the variation laws of collector voltage $V_1$, base thickness $W_p$, temperature difference $\Delta T$ across the device, and thermoelectric conversion efficiency $\eta$ with respect to emitter voltage $V_2$ and base carrier concentration $P_a$ under N type region doping concentration $N_{d1} = N_{d2} = 0.8 \times 10^{25}$ m$^{-3}$. From Figure 7 (a), under the same $P_a$, as the emitter voltage $V_2$ increases, the collector voltage $V_1$ increases. Additionally, as the base carrier concentration $P_a$ increases, the minimum and maximum allowable values of $V_2$ also increase, as the initial and maximum reverse bias voltages $V_1$ required to be applied to the collector. From Figure 7 (b), it can be observed that the base thickness $W_p$ decreases as the emitter voltage $V_2$ and base carrier concentration $P_a$ increase. From Figure 7 (c), when $P_a$ is constant, the temperature difference $\Delta T$ across the device decreases as $V_2$ increases. When $V_2$ is constant, $\Delta T$ increases with $P_a$, and the maximum temperature difference $\Delta T_{max}$ across the device also increases with $P_a$. In Figure 7(d), the thermoelectric conversion efficiency $\eta$ decreases as $P_a$ increases when $V_2$ is constant and increases as $V_2$ increases when $P_a$ is constant. Combining Figures 7 (a), (b), (c), and (d), it can be concluded that when $P_a$ is constant, the temperature difference $\Delta T$ across the device is maximized when $V_2$ is at its minimum allowable value, with corresponding minimum values of $V_1$ and $W_p$, and $\eta = 0$. Comparing $\Delta T_{max}$ under different $P_a$ conditions, the maximum temperature difference across the device increases from $\Delta T_{max} = 207.02$ K at $V_2 = 0.00156$ V, $V_1 = 0.1585$ V, and $P_a = 8 \times 10^{25}$ m$^{-3}$ to $\Delta T_{max} = 238.23$ K at $V_2 = 0.02656$ V, $V_1 = 0.2329$ V, and $P_a = 100 \times 10^{25}$ m$^{-3}$. At these conditions, the base thickness $W_p$ is 1.96 nm and 0.19 nm, respectively.

### 3.4 The calculation result of $N_{d1}=N_{d2}=0.9\times10^{25}$m$^{-3}$

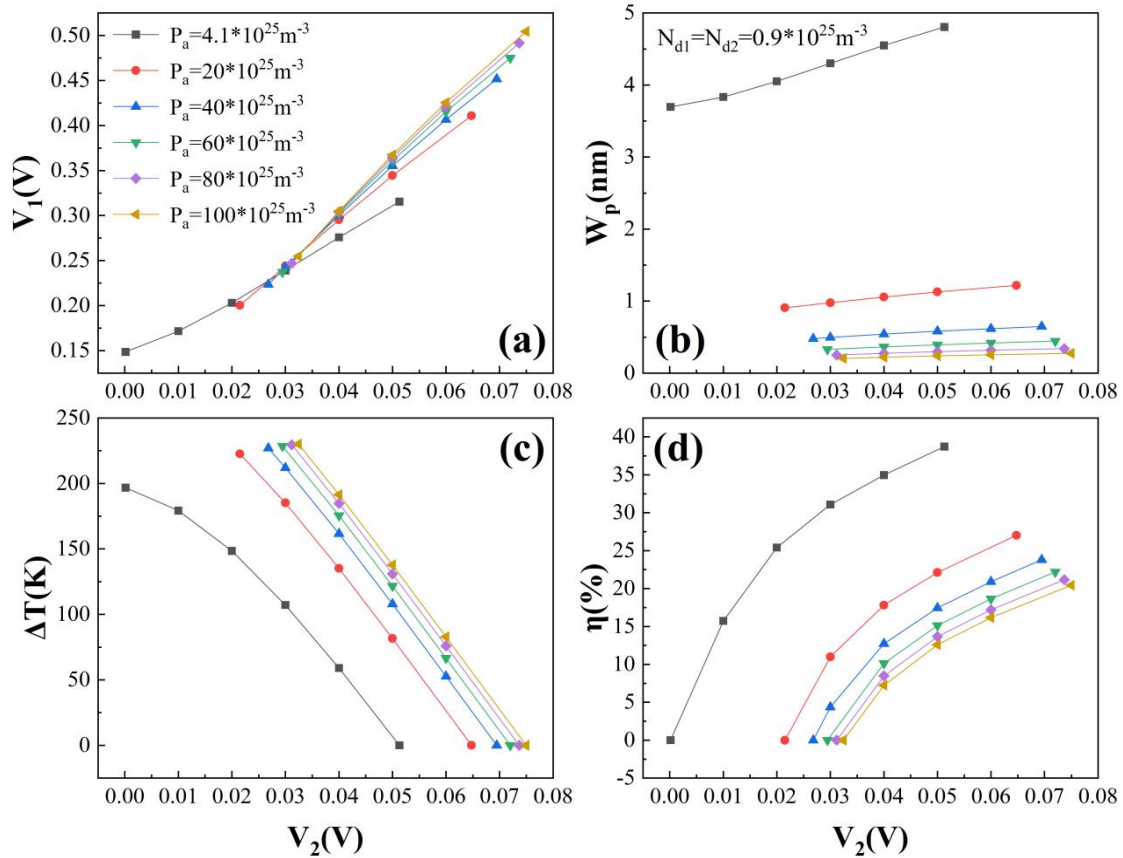

Figure 8 $V_1$ (a), $W_p$ (b), $\Delta T$ (c), $\eta$ (d) as functions of $V_2$ and $P_a$ when $N_d = 0.9 \times 10^{25}$ m$^{-3}$

Figure 8 (a), (b), (c), and (d) show the variation laws of collector voltage $V_1$, base thickness $W_p$, temperature difference $\Delta T$ across the device, and thermoelectric conversion efficiency $\eta$ with respect to emitter voltage $V_2$ and base carrier concentration $P_a$ under the condition of N type region doping concentration $N_{d1} = N_{d2} = 0.9 \times 10^{25}$ m$^{-3}$. From Figure 8 (a), under the same $P_a$, as the emitter voltage $V_2$ increases, the collector voltage $V_1$ also increases. Additionally, as the base carrier concentration $P_a$ increases, the minimum and maximum allowable values of $V_2$ increase, as do the initial and maximum reverse bias voltages $V_1$ that need to be applied to the collector. From Figure 8 (b), it can be observed that the base thickness $W_p$ decreases as the emitter voltage $V_2$ and base carrier concentration $P_a$ increases. From Figure 8 (c), when $P_a$ is constant, the temperature difference $\Delta T$ across the device decreases as $V_2$ increases. Conversely, when $V_2$ is constant, $\Delta T$ increases with $P_a$, and the maximum temperature difference $\Delta T_{max}$ also increases as $P_a$ increases. In Figure 8(d), the thermoelectric conversion efficiency η decreases as $P_a$ increases when $V_2$ is constant and increases as $V_2$ increases when $P_a$ is constant. Combining the information from Figures 8 (a), (b), (c), and (d), it can be concluded that when $P_a$ is constant, the temperature difference $\Delta T$ across the device is maximized when $V_2$ is at its minimum allowable value, with corresponding minimum values of $V_1$ and $W_p$, and η = 0. Comparing the maximum temperature differences $\Delta T_{max}$ under different $P_a$ conditions, the maximum temperature difference increases from $\Delta T_{max}$ = 196.75 K at $V_2$ = 0.000154 V, $V_1$ = 0.1485 V, and $P_a$ = 4.1×10²⁵ m⁻³ to $\Delta T_{max}$ = 230.12 K at $V_2$ = 0.03245 V, $V_1$ = 0.2548 V, and $P_a$ = 100×10²⁵ m⁻³. At these conditions, the base thickness $W_p$ is 3.70 nm and 0.20 nm, respectively.

## 3.5 The calculation result of $N_{d1}=N_{d2}=1.0\times10^{25}$ m⁻³

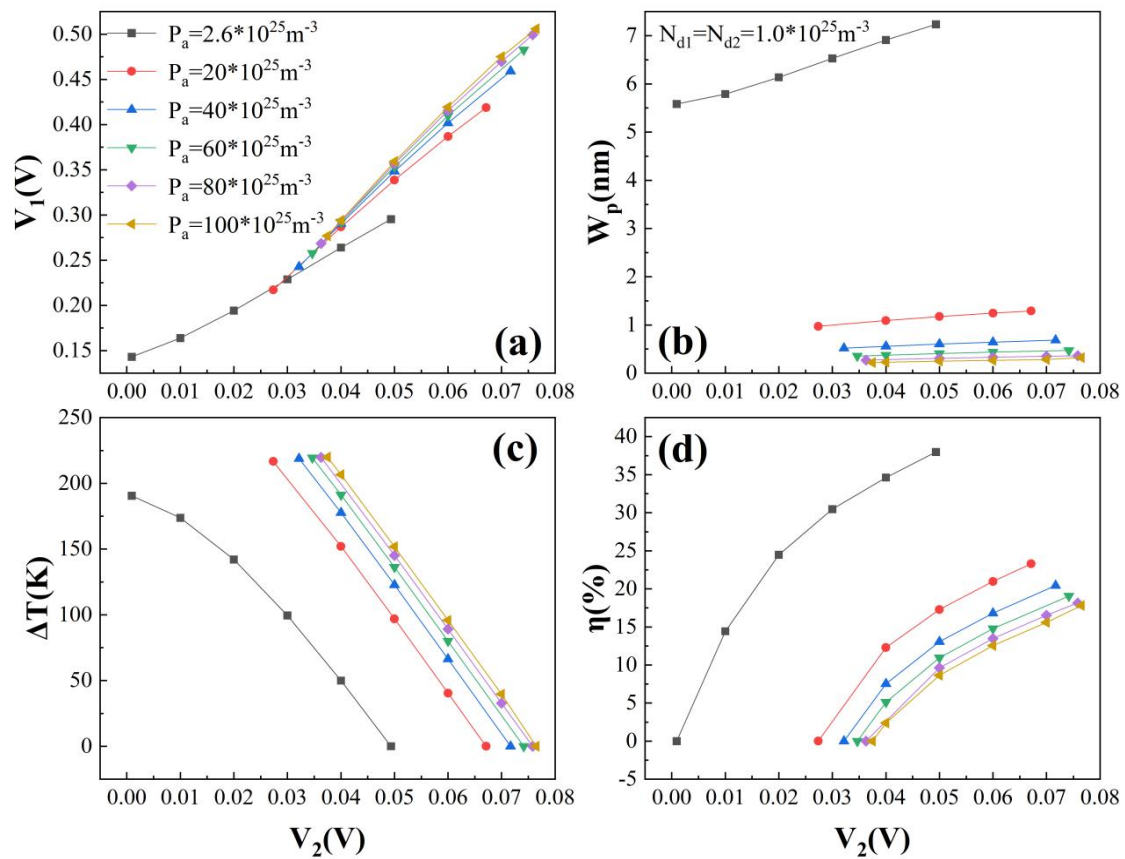

Figure 9 $V_1$ (a), $W_p$ (b), $\Delta T$ (c), η (d) as functions of $V_2$ and $P_a$ when $N_d$ = 1.0×10²⁵ m⁻³

Figure 9 (a), (b), (c), and (d) show the variation laws of collector voltage $V_1$, base thickness $W_p$, temperature difference $\Delta T$ across the device, and thermoelectric conversion efficiency η with respect to emitter voltage $V_2$ and base carrier concentration $P_a$ under the condition of N type region doping concentration $N_{d1}$ = $N_{d2}$ = 1.0×10²⁵ m⁻³. From Figure 9 (a), under the same $P_a$, as the emitter voltage $V_2$ increases, the collector voltage $V_1$ also increases. Additionally, as the base carrier concentration $P_a$ increases, the minimum and maximum allowable values of $V_2$ increase, as do the initial and maximum reverse bias voltages $V_1$ that need to be applied to the collector. From Figure 9 (b), it can be observed that the base thickness $W_p$ decreases as the emitter voltage $V_2$ and base carrier concentration $P_a$ increases. From Figure 9 (c), when $P_a$ is constant, the temperature difference $\Delta T$ across the device decreases as $V_2$ increases. Conversely, when $V_2$ is constant, $\Delta T$ increases with $P_a$, and the maximum temperature difference $\Delta T_{max}$ also increases as $P_a$ increases. In Figure 9 (d), the thermoelectric conversion efficiency $\eta$ decreases as $P_a$ increases when $V_2$ is constant and increases as $V_2$ increases when $P_a$ is constant. Combining the information from Figures 9 (a), (b), (c), and (d), it can be concluded that when $P_a$ is constant, the temperature difference $\Delta T$ across the device is maximized when $V_2$ is at its minimum allowable value, with corresponding minimum values of $V_1$ and $W_p$, and $\eta = 0$. Comparing the maximum temperature differences $\Delta T_{max}$ under different $P_a$ conditions, the maximum temperature difference increases from $\Delta T_{max} = 190.46$ K at $V_2 = 0.00092$ V, $V_1 = 0.1430$ V, and $P_a = 2.6\times10^{25}$ m$^{-3}$ to $\Delta T_{max} = 219.97$ K at $V_2 = 0.03750$ V, $V_1 = 0.2768$ V, and $P_a = 100\times10^{25}$ m$^{-3}$. At these conditions, the base thickness $W_p$ is 5.58 nm and 0.22 nm, respectively.

## 3.6 The calculation result of $N_{d1}=N_{d2}=1.5\times10^{25}$m$^{-3}$

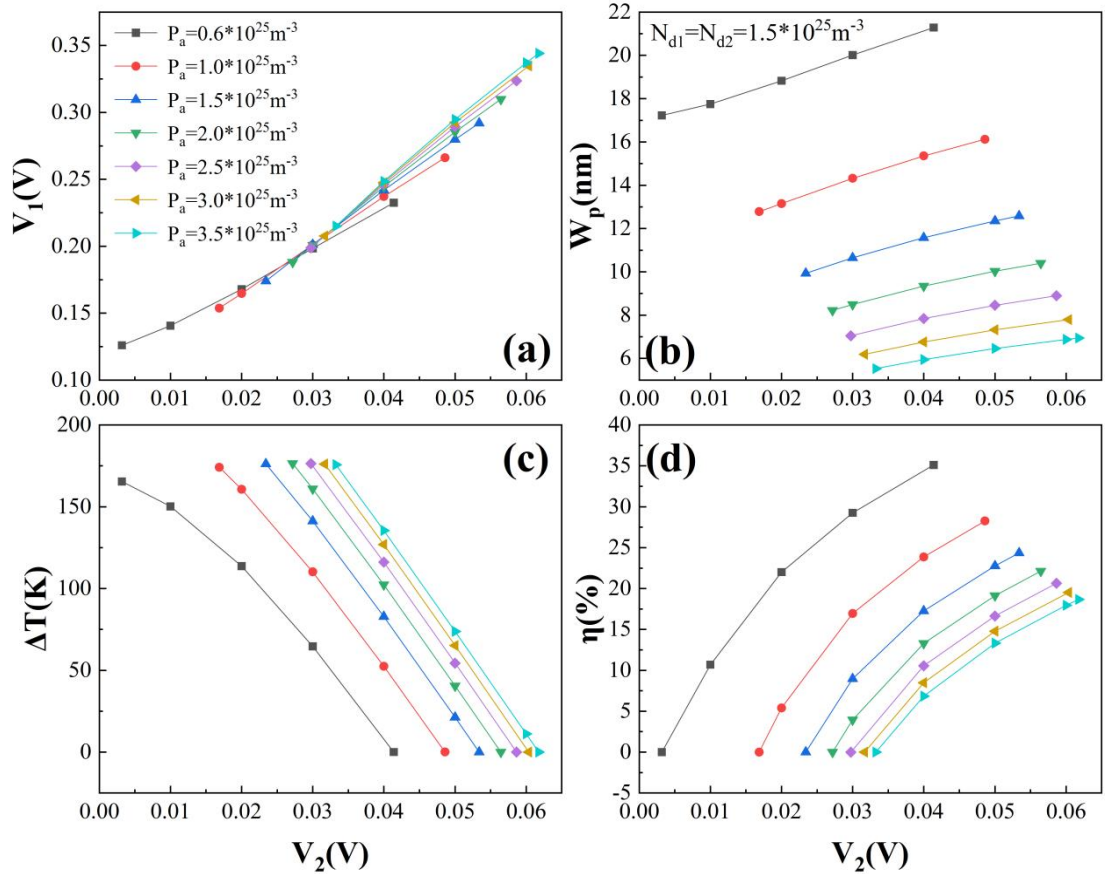

Figure 10 $V_1$ (a), $W_p$ (b), $\Delta T$ (c), $\eta$ (d) as functions of $V_2$ and $P_a$ when $N_d = 1.5\times10^{25}$ m$^{-3}$

Figure 10 (a), (b), (c), (d) show the variation laws of collector voltage $V_1$, base depletion layer thickness $W_p$, temperature difference $\Delta T$ across the device, and thermoelectric conversion efficiency η with respect to emitter voltage $V_2$ and base carrier concentration $P_a$ under the condition of N type region doping concentration $N_{d1} = N_{d2} = 1.5 \times 10^{25}$ m$^{-3}$. From Figure 10 (a), under the same $P_a$, as the emitter voltage $V_2$ increases, the collector voltage $V_1$ also increases. Additionally, as the base carrier concentration $P_a$ increases, the minimum and maximum allowable values of $V_2$, as well as the initial and maximum reverse bias voltages $V_1$ required to be applied to the collector, also increase. From Figure 10 (b), it can be observed that the base thickness $W_p$ decreases as the emitter voltage $V_2$ and the base carrier concentration $P_a$ increases. From Figure 10 (c), when Pa $P_a$ is constant, the temperature difference ΔT across the device decreases with increasing $V_2$. When $V_2$ is constant, ΔT increases with increasing $P_a$. Furthermore, the maximum temperature difference ΔT$_{max}$ across the device first increases and then decreases as $P_a$ increases, reaching its maximum value when $P_a = 2.0 \times 10^{25}$ m$^{-3}$. From the comprehensive analysis of Figures 10 (a), (b), (c), and (d), it can be concluded that when $P_a$ is constant, the temperature difference ΔT across the device is maximized when $V_2$ is at its minimum allowable value, with corresponding minimum values of $V_1$ and $W_p$, and η = 0. By comparing the maximum temperature differences ΔT$_{max}$ under different $P_a$ conditions, the maximum temperature difference across the device increases from ΔT$_{max}$ = 165.44 K at $V_2$ = 0.00318 V, $V_1$ = 0.1260 V, and $P_a = 0.6 \times 10^{25}$ m$^{-3}$, to ΔT$_{max}$ = 176.48 K at $V_2$ = 0.02717 V, $V_1$ = 0.1880 V, and $P_a = 2.0 \times 10^{25}$ m$^{-3}$, and then decreases to ΔT$_{max}$ = 175.75 K at $V_2$ = 0.03325 V, $V_1$ = 0.2150 V, and $P_a = 3.5 \times 10^{25}$ m$^{-3}$. The corresponding base thicknesses $W_p$ are 17.23 nm, 8.22 nm, and 5.53 nm, respectively.

### 3.7 The calculation result of $N_{d1}=N_{d2}=2.0 \times 10^{25}$m$^{-3}$

Figure 11 (a), (b), (c), and (d) show the variation laws of collector voltage $V_1$, base region depletion layer thickness $W_p$, temperature difference ΔT across the device, and thermoelectric conversion efficiency η with respect to emitter voltage $V_2$ and base carrier concentration $P_a$, under the condition of N type region doping concentration $N_{d1} = N_{d2} = 2.0 \times 10^{25}$ m$^{-3}$. From Figure 11(a), it can be observed that under the same $P_a$, as the voltage $V_2$ applied to the emitter increases, the collector biot voltage $V_1$ also increases. Additionally, as the base carrier concentration $P_a$ increases, the minimum and maximum allowable values of $V_2$, as well as the initial and maximum reverse bias voltages $V_1$ required to be applied to the collector, also increase. Furthermore, compared with Figures 5 to 10, it can be found that as $N_d$ increases, the initial allowable value of $V_1$ increases, and the numerical range decreases rapidly. From Figure 11 (b), the base region thickness $W_p$ decreases as the voltage $V_2$ applied to the emitter and the base carrier concentration $P_a$ increases. From Figure 11 (c), when Pa is constant, it can be observed that as $V_2$ increases, the temperature difference ΔT across the device decreases. When $V_2$ is constant, as $P_a$ increases, ΔT increases. Moreover, the maximum temperature difference ΔTmax across the device first increases and then decreases as $P_a$ increases, reaching its maximum value when $P_a = 0.4 \times 10^{25}$ m$^{-3}$, with ΔT$_{max}$ = 149.56 K. From Figure 11 (d), the thermoelectric conversion efficiency η decreases as $P_a$ increases when $V_2$ is constant and increases as $V_2$ increases when $P_a$ is constant. Integrating the information from Figures 11 (a), (b), (c), and (d), it can be concluded that when $P_a$ is constant, the temperature difference ΔT across the device is maximized when $V_2$ is at its minimum allowable value, with corresponding minimum values of $V_1$ and $W_p$, and η = 0. Comparing the maximum temperature differences ΔT$_{max}$ under different $P_a$ conditions, the maximum temperature difference across the device increases from ΔT$_{max}$ = 149.56 K at $V_2$ = 0.00962 V, $V_1$ = 0.1263 V, and $P_a = 0.3 \times 10^{25}$ m$^{-3}$, to ΔT$_{max}$ = 151.05 K at $V_2$ = 0.01806 V, $V_1$ = 0.1452 V, and $P_a = 0.4 \times 10^{25}$ m$^{-3}$, and then decreases to $\Delta T_{max} = 133.12$ K at $V_2 = 0.04699$ V, $V_1 = 0.2697$ V, and $P_a = 3.5\times10^{25}$ m$^{-3}$. The corresponding base region thicknesses $W_p$ are 26.33 nm, 23.20 nm, and 6.74 nm, respectively.

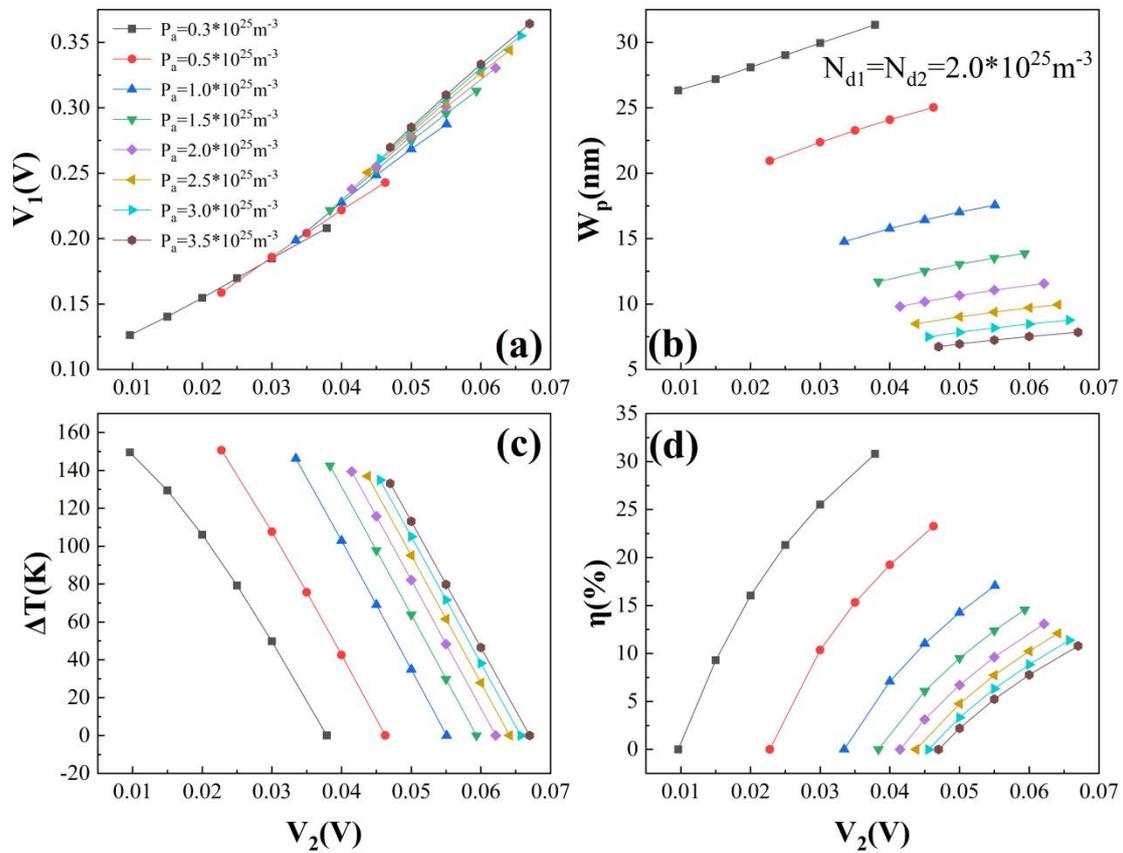

Figure 11 $V_1$ (a), $W_p$ (b), $\Delta T$ (c), $\eta$ (d) as functions of $V_2$ and $P_a$ when $N_d = 2.0\times10^{25}$ m$^{-3}$

## 3.8 The calculation result of $N_{d1}=N_{d2}=2.5\times10^{25}$m$^{-3}$

Figure 12 (a), (b), (c), and (d) show the variation laws of collector voltage $V_1$, base region thickness $W_p$, temperature difference $\Delta T$ across the device, and thermoelectric conversion efficiency $\eta$ with respect to emitter voltage $V_2$ and base carrier concentration $P_a$ under the condition of N type region doping concentration $N_{d1} = N_{d2} = 2.5\times10^{25}$ m$^{-3}$. From Figure 12 (a), under the same $P_a$, as the emitter voltage $V_2$ increases, the collector voltage $V_1$ also increases. Additionally, as the base carrier concentration $P_a$ increases, the minimum and maximum allowable values of $V_2$ increase, as do the initial and maximum reverse bias voltages $V_1$ that need to be applied to the collector. Comparing with Figures 5 to 11, it can be observed that as $N_d$ increases, the initial allowable value of $V_1$ further increases, and its numerical range further decreases. From Figure 12 (b), the base region thickness $W_p$ decreases as the emitter voltage $V_2$ and base carrier concentration $P_a$ increases. From Figure 12 (c), when $P_a$ is constant, it can be observed that as $V_2$ increases, the temperature difference $\Delta T$ across the device decreases. Conversely, when $V_2$ is constant, as $P_a$ increases, $\Delta T$ increases. From the comprehensive analysis of Figures 12 (a), (b), (c) and (d), it can be observed that when $P_a$ is constant, $V_2$ attains its minimum allowable value, resulting in the maximum temperature difference $\Delta T$ across the device terminals. Correspondingly, $V_1$ and $W_p$ also reach their minimum values, with $\eta=0$. By comparing $\Delta T_{max}$ under different $P_a$ conditions, the maximum temperature difference across the device terminals increases from

ΔT$_{max}$=133.90K at V$_2$=0.00569V, V$_1$=0.1136V, and P$_a$=0.16×10$^{25}$m$^{-3}$, to ΔT$_{max}$=134.54K at V$_2$=0.01168V, V$_1$=0.1241V, and P$_a$=0.18×10$^{25}$m$^{-3}$, and then decreases to ΔT$_{max}$=89.92K at V$_2$=0.05743V, V$_1$=0.3165V, and P$_a$=3.5×10$^{25}$m$^{-3}$. The corresponding base region thicknesses W$_p$ are 36.06nm, 34.54nm, and 7.81nm, respectively.

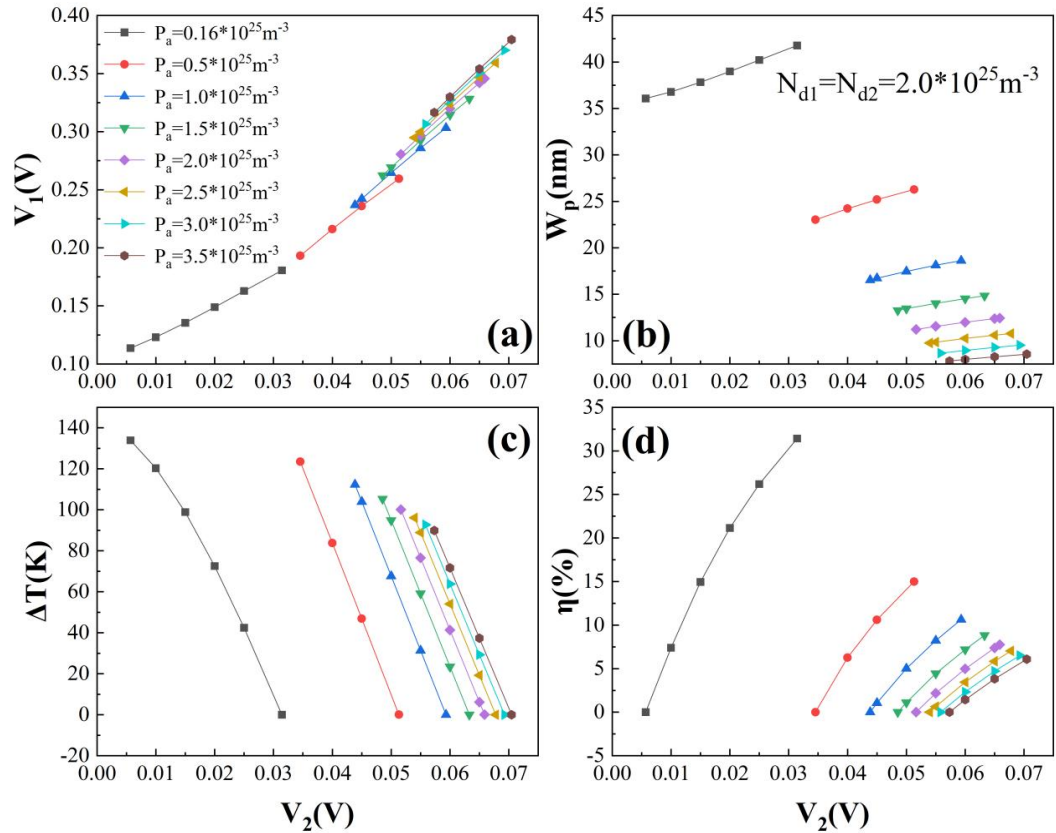

Figure 12 V$_1$ (a), W$_p$ (b), ΔT (c), η (d) as functions of V$_2$ and P$_a$ when N$_d$ = 2.5×10$^{25}$ m$^{-3}$

### 3.9 The calculation result of N$_{d1}$=N$_{d2}$=3.0×10$^{25}$m$^{-3}$

Figure 13 (a), (b), (c), and (d) show the variation laws of collector voltage V$_1$, base thickness W$_p$, temperature difference ΔT across the device, and thermoelectric conversion efficiency η with respect to emitter voltage V$_2$ and base carrier concentration P$_a$ under N type region doping concentrations N$_{d1}$ = N$_{d2}$ = 3.0×10$^{25}$ m$^{-3}$. From Figure 13 (a), under the same P$_a$, as the voltage V$_2$ applied to the emitter increases, the collector voltage V$_1$ increases. Additionally, as the base carrier concentration P$_a$ increases, the minimum and maximum allowable values of V$_2$ also increase, as do the initial and maximum reverse bias voltages V$_1$ that need to be applied to the collector. Comparing with Figures 5 to 12, it can be observed that as N$_d$ increases, the initial allowable value of V$_1$ further increases, and its numerical range further decreases. From Figure 13 (b), the base thickness W$_p$ decreases as the voltage V$_2$ applied to the emitter and the base carrier concentration P$_a$ increase. From Figure 13 (c), it can be observed that when P$_a$ is constant, the temperature difference ΔT across the device decreases as V$_2$ increases. When V$_2$ is constant, ΔT increases as P$_a$ increases, and the maximum temperature difference ΔT$_{max}$ across the device decreases as P$_a$ increases. From Figure 13 (d), the thermoelectric conversion efficiency η decreases as P$_a$ increases when V$_2$ is constant and increases as V$_2$ increases when P$_a$ is constant. Combining Figures 13 (a),

(b), (c), and (d), it can be concluded that when $P_a$ is constant, the minimum allowable value of $V_2$ is taken, resulting in the maximum temperature difference $\Delta T$ across the device, with the corresponding $V_1$ and $W_p$ also taking their minimum values, and $\eta = 0$. Comparing the maximum temperature differences $\Delta T_{max}$ under different $P_a$ conditions, the maximum temperature difference across the device decreases from $\Delta T_{max} = 122.95$ K at $V_2 = 0.00613$ V, $V_1 = 0.1107$ V, and $P_a = 0.11\times10^{25}$ m$^{-3}$ to $\Delta T_{max} = 49.69$ K at $V_2 = 0.06584$ V, $V_1 = 0.3568$ V, and $P_a = 3.5\times10^{25}$ m$^{-3}$. The corresponding base thicknesses $W_p$ are 42.26 nm and 8.73 nm, respectively.

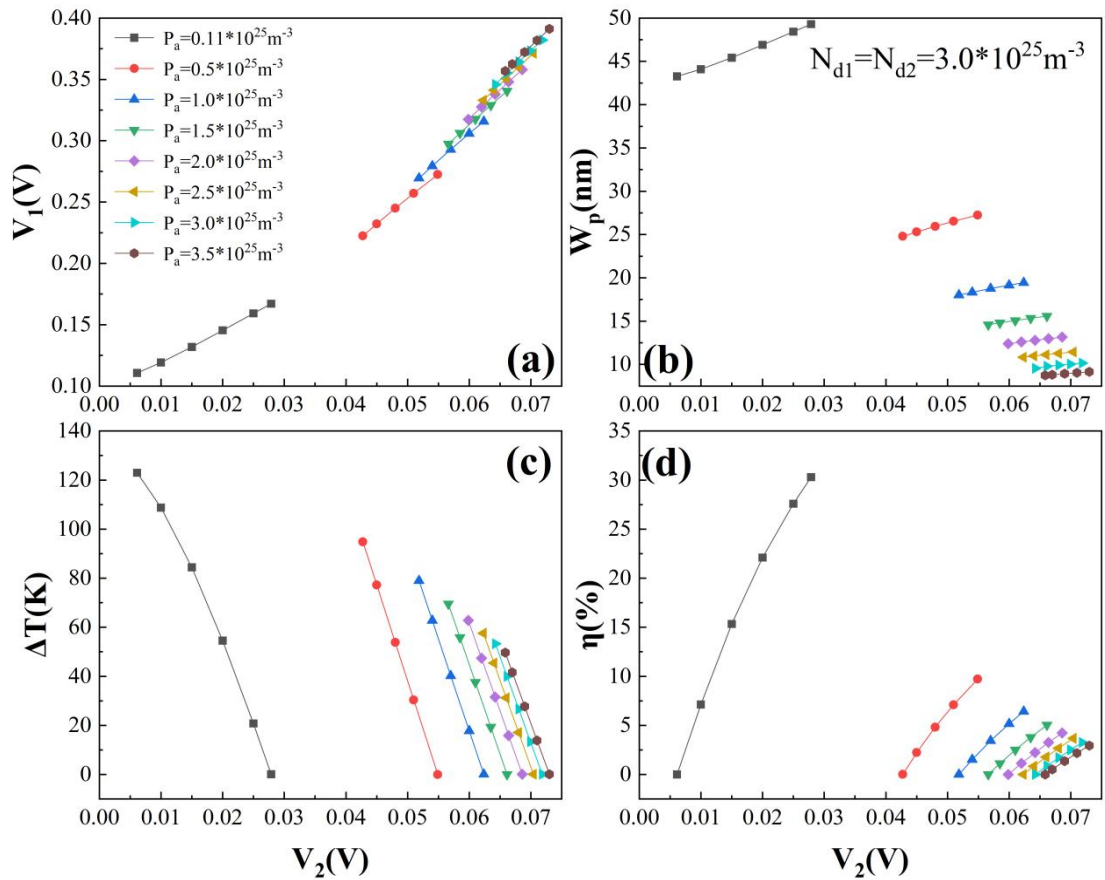

Figure 13 $V_1$ (a), $W_p$ (b), $\Delta T$ (c), $\eta$ (d) as functions of $V_2$ and $P_a$ when $N_d = 3.0\times10^{25}$ m$^{-3}$

### 3.10 The calculation result of $N_{d1}=N_{d2}=3.5\times10^{25}$m$^{-3}$

Figure 14 (a), (b), (c), and (d) show the variation laws of collector voltage $V_1$, base thickness $W_p$, temperature difference $\Delta T$ across the device, and thermoelectric conversion efficiency $\eta$ with respect to emitter voltage $V_2$ and base carrier concentration $P_a$ under N type region doping concentrations $N_{d1} = N_{d2} = 3.5\times10^{25}$ m$^{-3}$. From Figure 14 (a), under the same $P_a$, as the emitter voltage $V_2$ increases, the collector voltage $V_1$ also increases. Additionally, as the base carrier concentration $P_a$ increases, the minimum and maximum allowable values of $V_2$, as well as the starting and maximum reverse bias voltages $V_1$ required to be applied to the collector, also increase. Comparing with Figures 5 to 13, it can be observed that as $N_d$ increases, the starting value of allowable $V_1$ further increases, and its numerical range further decreases. From Figure 14 (b), the base thickness $W_p$ decreases as the emitter voltage $V_2$ and base carrier concentration $P_a$ increases. From Figure 14 (c), it can be observed that when $P_a$ is constant, the temperature difference $\Delta T$ across the device decreases as $V_2$ increases; when $V_2$ is constant, $\Delta T$ increases as $P_a$ increases, and the maximum temperature difference $\Delta T_{max}$ across the device decreases as $P_a$ increases. In Figure 14 (d), the thermoelectric conversion efficiency $\eta$ decreases as $P_a$ increases when $V_2$ is constant and increases as $V_2$ increases when $P_a$ is constant. From the comprehensive analysis of Figures 14 (a), (b), (c), and (d), it can be seen that when $P_a$ is constant, the temperature difference $\Delta T$ across the device is maximized when $V_2$ is set to its minimum allowable value, and the corresponding $V_1$ and $W_p$ are also minimized, with $\eta = 0$. By comparing the maximum temperature differences $\Delta T_{max}$ under different $P_a$ conditions, the maximum temperature difference across the device decreases from $\Delta T_{max} = 114.32$ K at $V_2 = 0.00295$ V, $V_1 = 0.1038$ V, and $P_a = 0.08 \times 10^{25}$ m$^{-3}$ to $\Delta T_{max} = 12.78$ K at $V_2 = 0.07312$ V, $V_1 = 0.3924$ V, and $P_a = 3.5 \times 10^{25}$ m$^{-3}$. The corresponding base thicknesses $W_p$ are 50.05 nm and 9.53 nm, respectively.

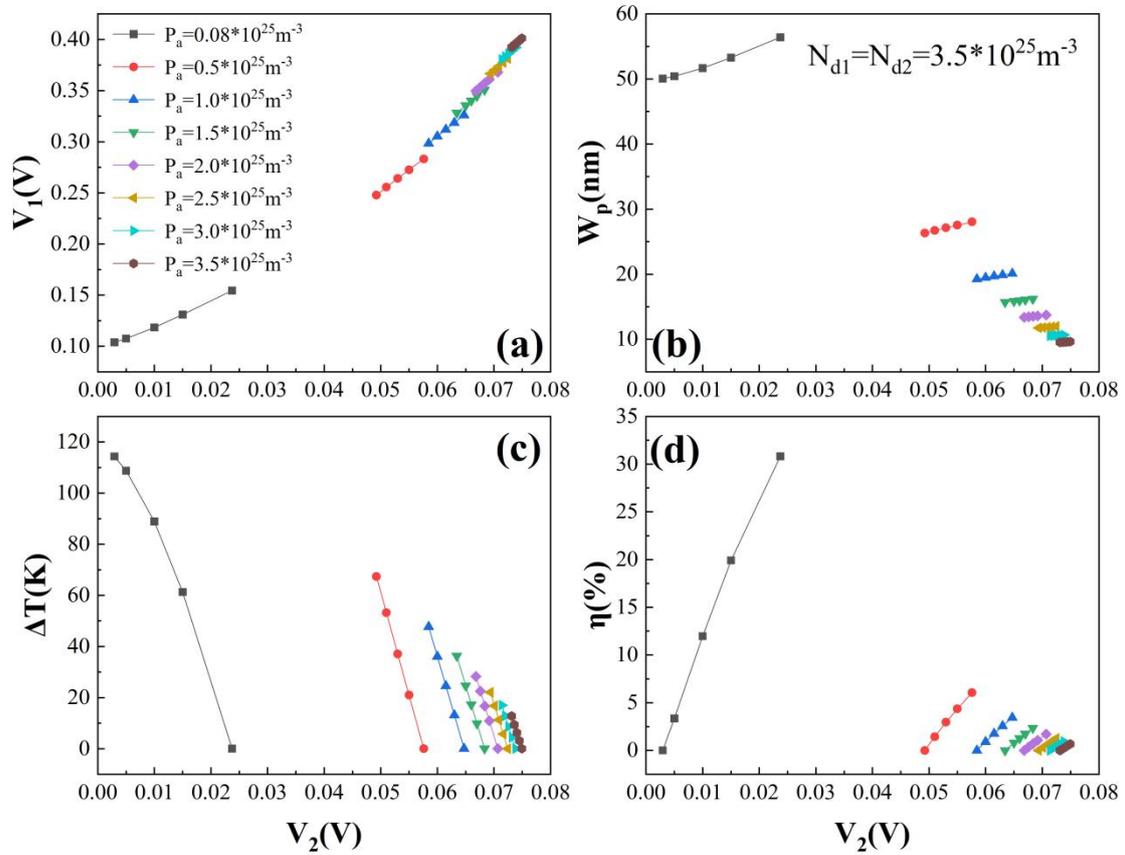

Figure 14 $V_1$ (a), $W_p$ (b), $\Delta T$ (c), $\eta$ (d) as functions of $V_2$ and $P_a$ when $N_d = 3.5 \times 10^{25}$ m$^{-3}$

## 3.11 Comprehensive Data Analysis

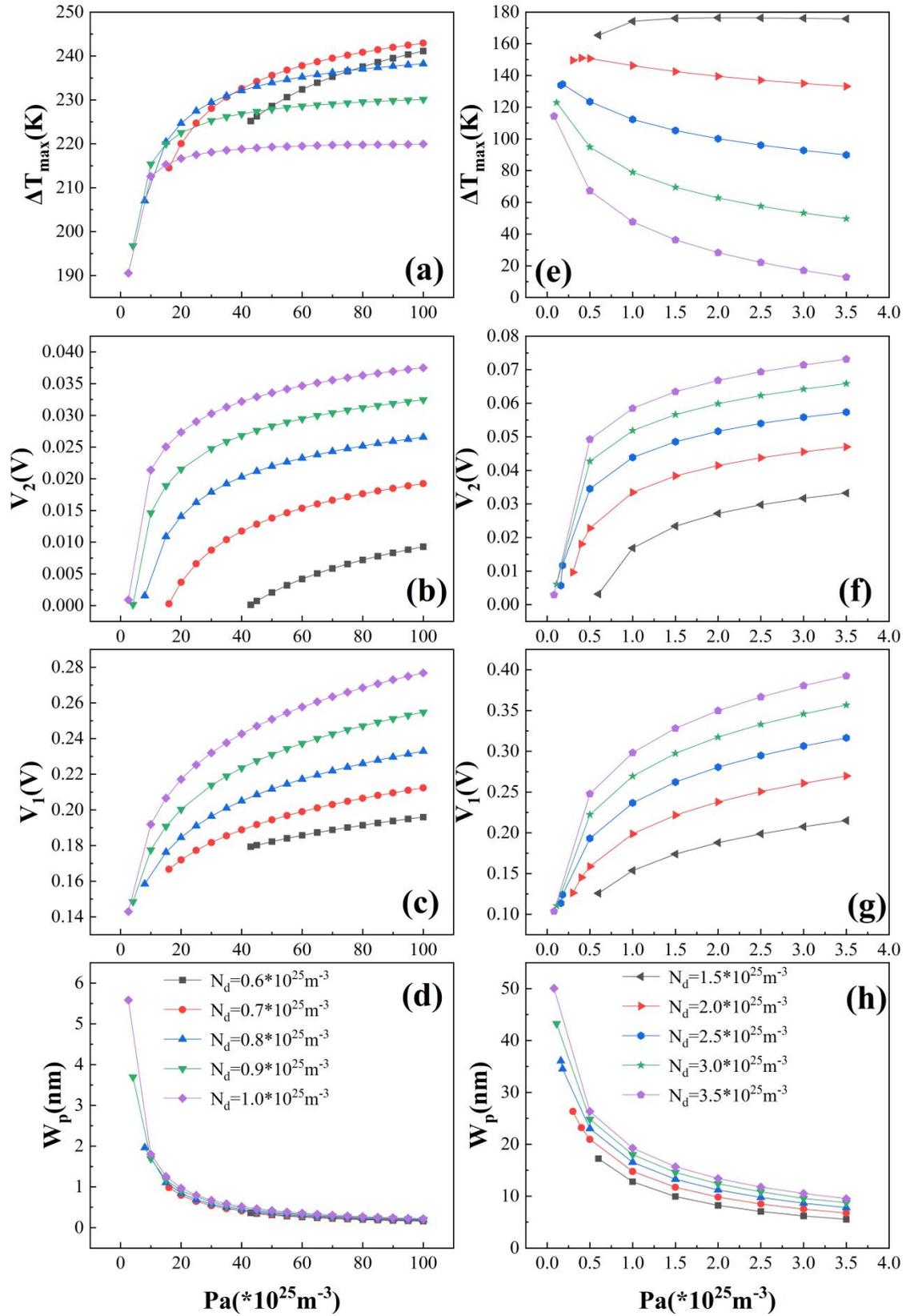

Figure 15 $\Delta T_{max}$ (a), $V_2$ (b), $V_1$ (c), and $W_p$ (d) versus base region concentration $P_a$ when $N_d$ = 0.6–1.0 × 10$^{25}$ m$^{-3}$, and $\Delta T_{max}$ (e), $V_2$ (f), $V_1$ (g), and $W_p$ (h) versus base region concentration $P_a$ when $N_d$=1.5–3.5 × 10$^{25}$ m$^{-3}$

Figure 15 shows $\Delta T_{max}$, $V_1$, $V_2$, and $W_p$ versus base region concentration $P_a$ when the N type region doping concentration $N_d$ is 0.6-3.5×10$^{25}$m$^{-3}$. As shown in Figure 15 (e), when $N_d$ is 1.5-3.5×10$^{25}$m$^{-3}$, the maximum value of $\Delta T_{max}$ has already appeared in the range of $P_a$ = 0.08-2.1×10$^{25}$m$^{-3}$. Therefore, in Figures 15 (e) to (h), only the calculation results for $P_a \le$ 3.5×10$^{25}$m$^{-3}$ are provided. In contrast, in Figures 15 (a) to (d), the calculation results are presented when the value of $P_a$ ranging from 4.1 to 100×10$^{25}$m$^{-3}$.

From Figure 15 (a), when the N type region doping concentration $N_d$ is in the range of 0.6-1.0×10$^{25}$ m$^{-3}$, the N type region doping concentration $N_d$ increases, the minimum permissible base region carrier concentration $P_{amin}$ decreases. For any given $N_d$ value, $\Delta T_{max}$ increases with the increase of $P_a$, Moreover, as $P_a$ increases, $\Delta T_{max}$ increases rapidly at first, reaches a certain inflection point, and then increases slowly or even tends to stabilize. The maximum value of $\Delta T_{max}$ corresponding to all 5 $N_d$ values is obtained at $P_a$ = 100×10$^{25}$ m$^{-3}$. Besides changing with $P_a$, $\Delta T_{max}$ also changes with $N_d$, that is, $\Delta T_{max}$ changes simultaneously with both $P_a$ and $N_d$. When $2.6 \le P_a <$ 4.1×10$^{25}$ m$^{-3}$, the maximum temperature difference $\Delta T_{max}$ reaches its maximum value of 190.53 - 196.45 K when $N_d$ = 1.0×10$^{25}$ m$^{-3}$, and the corresponding $W_p$ is 5.58–3.92 nm. When $4.1 \le P_a <$ 20×10$^{25}$ m$^{-3}$, $\Delta T_{max}$ reaches its maximum value of 196.75–222.53 K when $N_d$ = 0.9×10$^{25}$ m$^{-3}$, and the corresponding $W_p$ is 3.70–0.91 nm. When $20 \le P_a <$ 40×10$^{25}$ m$^{-3}$, $\Delta T_{max}$ reaches its maximum value of 224.71–232.13 K when Nd = 0.8×10$^{25}$ m$^{-3}$, and the corresponding $W_p$ is 0.84–0.44 nm When $40 \le P_a \le$ 100×10$^{25}$ m$^{-3}$, $\Delta T_{max}$ reaches its maximum value of 232.55–242.89 K when $N_d$ = 0.7×10$^{25}$ m$^{-3}$, and the corresponding $W_p$ is 0.41–0.17 nm. That is, as the N type region doping concentration Nd increases, $\Delta T_{max}$ first increases, reaches a maximum, and then decreases. For example, starting from $\Delta T_{max}$ = 241.10 K at $N_d$ = 0.6×10$^{25}$ m$^{-3}$ and $P_a$ = 100×10$^{25}$ m$^{-3}$, $\Delta T_{max}$ increases, reaching its maximum of 242.93 K at $N_d$ = 0.7×10$^{25}$ m$^{-3}$ and $P_a$ = 100×10$^{25}$ m$^{-3}$, and then begins to decrease, such as reaching a maximum of 219.97 K at $N_d$ = 1.0×10$^{25}$ m$^{-3}$ and $P_a$ = 100×10$^{25}$ m$^{-3}$. The maximum values of $\Delta T_{max}$ for N type region doping concentrations of 0.6, 0.7, 0.8, 0.9, and 1.0×10$^{25}$ m$^{-3}$ are 241.12 K, 242.93 K, 238.23 K, 230.12 K, and 219.97 K, respectively. That is, the maximum temperature difference $\Delta T_{max}$ = 242.93 K is achieved at $N_{d1}$ = $N_{d2}$ = 0.7×10$^{25}$ m$^{-3}$ and $P_a$ = 100×10$^{25}$ m$^{-3}$, which is also the maximum temperature difference obtained in this study. Corresponding to $\Delta T_{max}$ = 242.93K, $V_1$ = 0.2124 V, $V_2$ = 0.01924 V, and $W_p$ = 0.17 nm.

From Figure 15 (e), when the N type region doping concentration $N_d$ is in the range of 1.5-3.5×10$^{25}$ m$^{-3}$, the $\Delta T_{max}$ decreases with the increasing of $N_d$, and the corresponding $P_a$ value also decreases. Therefore, there is no calculation for $\Delta T$ under the condition of $N_d > 3.5×10^{25}$ m$^{-3}$. Additionally, for different $N_d$ values, the law of $\Delta T_{max}$ changing with $P_a$ is different. When the N type region doping concentration $N_d$ is in the range of 1.5-2.5×10$^{25}$ m$^{-3}$, $\Delta T_{max}$ varies with $P_a$ in a parabolic pattern. As $P_a$ increases, $\Delta T_{max}$ first increases, reaches its maximum at the vertex of the parabola, and then gradually decreases. Moreover, the maximum value of $\Delta T_{max}$ decreases as the doping concentration $N_d$ in the N region increases. When $N_d$ is $1.5 \times 10^{25}$m$^{-3}$, and the maximum value of $\Delta T_{max}$ is 176.48K and the corresponding parameters are $P_a = 2 \times 10^{25}$ m$^{-3}$, $V_1$ = 0.1880 V, $V_2$ = 0.02717 V, and $W_p$ = 8.22 nm. When the N type region doping concentration $N_d$ increases to 2.0×10$^{25}$ m$^{-3}$, the extremum point of $\Delta T_{max}$ shifts to $P_a$ = 0.4×10$^{25}$ m$^{-3}$, achieving a value of 151.04 K, with corresponding parameters $V_1$ = 0.1452 V, $V_2$ = 0.01806 V, and $W_p$ = 23.20 nm. When $N_d$ = 2.5×10$^{25}$ m$^{-3}$, the extremum point of $\Delta$Tmax occurs at $P_a$ = 0.18×10$^{25}$ m$^{-3}$, reaching 134.54 K, with $V_1$ = 0.1241 V, $V_2$ = 0.01168 V, and $W_p$ = 34.54 nm. For N type region doping concentrations $N_d \ge$

$3.0\times10^{25}$ m$^{-3}$, $\Delta T_{max}$ decreases as $P_a$ increases, that is, $\Delta T_{max}$ attains its maximum value at the minimum value of $P_a$. When $N_d = 3.0\times10^{25}$ m$^{-3}$, the maximum value of $\Delta T_{max}$ is 122.95 K, with $P_a = 0.11\times10^{25}$ m$^{-3}$, $V_1 = 0.1107$ V, $V_2 = 0.00613$ V, and $W_p = 43.26$ nm. When $N_d = 3.5\times10^{25}$ m$^{-3}$, the maximum value of $\Delta T_{max}$ is 114.32 K, and the corresponding parameters are $P_a = 0.08\times10^{25}$ m$^{-3}$, $V_1 = 0.1038$ V, $V_2 = 0.00295$ V, and $W_p = 50.05$ nm.

From Figure 15 (b), (c), (f), and (g), the externally applied voltages $V_2$ and $V_1$ at the emitter and collector increase with the base thickness and base concentration $P_a$, and also increase with the emitter and collector concentrations $N_d$. The reason is that according to equations (19) and (22), the interface voltage of the PN junction increases as carrier concentration increases, thus requiring a larger external voltage. From Figure 15 (d) and (h), it can be observed that the base thickness decreases with increasing base concentration $P_a$ and increases with increasing emitter and collector concentrations $N_d$, which aligns with equations (36) and (37).

## 4 Conclusion

In this study, an NPN junction thermoelectric transistor cooling device model composed of N-type $Bi_2Te_{2.97}Se_{0.03}$ and P-type $Bi_{0.5}Sb_{1.5}Te_3$ is constructed. By separately regulating the externally applied voltages on the emitter and collector, and coupling the transistor effect, thermoelectric effect, and interface effect within the NPN heterostructure, the temperature at the forward-biased end (emitter) increases while releasing heat, and the temperature at the reverse-biased end (collector) decreases while absorbing heat, thereby forming a thermoelectric transistor refrigerator and enhancing its cooling performance. Using the equivalent circuit method to convert the PNP heterostructure into a common-base circuit of a junction thermoelectric transistor, When the doping concentration in the N region is $N_d = 0.6$-$3.5\times10^{25}$ m$^{-3}$ and the doping concentration in the P region is $P_a = 0$-$100\times10^{25}$ m$^{-3}$, the effects of different applied voltages on the maximum temperature difference between the two ends of the thermoelectric transistor device and the base region thickness are investigated. The results are as follows:

(1) Under different N type region doping concentrations $N_d$ and base (P type region) doping concentrations $P_a$, as the voltage $V_2$ applied to the emitter increases, the collector voltage $V_1$, the base depletion layer thickness $W_p$, and the thermoelectric conversion efficiency $\eta$ all increase, while the temperature difference $\Delta T$ between the two ends of the thermoelectric transistor decreases;

(2) When the N type region doping concentration $N_d$ is in the range of 0.6-$1.0\times10^{25}$ m$^{-3}$, $\Delta T_{max}$ increases rapidly at first with increasing $P_a$, reaches a certain inflection point, then increases slowly or even tends to stabilize. As the N type region doping concentration $N_d$ increases, $\Delta T_{max}$ first increases, reaches a maximum at $N_d = 0.7\times10^{25}$ m$^{-3}$, and then decreases. When $N_d$ is greater than $1.5\times10^{25}$ m$^{-3}$, $\Delta T_{max}$ decreases as $N_d$ increases. When $N_d$ increases from $1.5\times10^{25}$ m$^{-3}$ to $3.5\times10^{25}$ m$^{-3}$, $\Delta T_{max}$ decreases from 176.51 to 114.32 K. When $N_d$ is in the range of 1.5-$2.5\times10^{25}$ m$^{-3}$, $\Delta T_{max}$ varies with $P_a$ in a parabolic pattern, where $\Delta T_{max}$ increases with increasing $P_a$, reaches its maximum at the vertex of the parabola, and then decreases slowly. When $N_d \geq 3.0\times10^{25}$ m$^{-3}$, $\Delta T_{max}$ decreases as $P_a$ increases, the maximum value of $\Delta T_{max}$ is obtained at the minimum value of $P_a$. Overall, as the N type region doping concentration $N_d$ increases, the minimum allowable base carrier concentration $P_{amin}$ decreases.

(3) The maximum temperature difference $\Delta T_{max}$ obtained is $\Delta T_{max} = 242.89$ K under the conditions of emitter voltage $V_2 = 0.01925$ V, collector voltage $V_1 = 0.2124$ V, P type region carrier concentration $P_a = 100\times10^{25}$ m$^{-3}$, and N type region carrier concentration $N_d = 0.7\times10^{25}$ m$^{-3}$. However, the corresponding base width $W_p = 0.17$ nm is not only difficult to achieve but also poses a risk of breakdown due to its extremely small size. Therefore, in actual production, suitable doping concentrations and external voltages need to be selected based on the actual process and the results of the optimized design. Even if the base region width is limited to the ?nm level, when the carrier concentration in the N region ($N_d$) is $1.5\times10^{25}$ m$^{-3}$, the carrier concentration in the base region ($P_a$) is $1.0\times10^{25}$ m$^{-3}$, the emitter voltage $V_2$ is 0.01686V, the collector voltage $V_1$ is 0.1838V, and the base region width ($W_p$) is 12.78nm, the maximum temperature difference $\Delta T_{max}$ that can be obtained is approximately 174.15 K. Therefore, the thermoelectric transistor device, which combines the thermoelectric effect with transistor technology can effectively enhance the maximum temperature difference that thermoelectric cooling technology can achieve.

Acknowledgments:
This work was financially supported by The National Key Research and Development Program of China [Grand No. 2017YFF0204706] and the Fundamental Research Funds for the Central Universities [Grant No. FRF-MP-18-005 and FRF-MP-19-005].

Reference
[1]  PELTIER J. Nouvelles experiences sur la caloricite des courants electrique[J]. Annales de Chimie et de Physique, 1834,56: 371-386.
[2]  J. Mao, G. Chen, Z. Ren. Thermoelectric cooling materials. Nat. Mater. 2021, 20, 454.
[3]  J. Yang, G. Li, H. Zhu, N. Chen, T. Lu, J. Gao, L. Guo, J. Xiang, P. Sun, Y. Yao, R. Yang, H. Zhao. Next-generation thermoelectric cooling modules based on high-performance Mg$_3$(Bi, Sb)$_2$ material. Joule 2022, 6, 193.
[4]  Y. Xiao, H. Wu, H. Shi, L. Xu, Y. Zhu, Y. Qin, G. Peng, Y. Zhang, Z. H. Ge, X. Ding, L. D. Zhao. High-Ranged ZT Value Promotes Thermoelectric Cooling and Power Generation in n-Type PbTe. Adv. En-ergy Mater. 2022.
[5]  Sun J-C, Su X-L, Yan Y-G, Liu W, Tan G-J, Tang X-F. Enhancing thermoelectric performance of n-type PbSe through forming solid solution with PbTe and PbS. ACS Appl. Energy Mater. 2020; 3: 2-8 (2020).
[6]  Perumal S, Samanta M, Ghosh T, Shenoy U-S, Bohra AK, Bhattacharya S, Singh A, Waghmare U V, Biswas K. Realization of high thermoelectric figure of merit in GeTe by complementary Co-doping of Bi and In. Joule. 2019; 3: 2565-80.
[7]  G.D. Mahan Introduction to thermoelectrics. APL Mater. 4 (2016), 104806.
[8]  J.T. Wei, L.L. Yang, Z. Ma, P.S. Song, M.L. Zhang, J. Ma, F.H. Yang, X.D. Wang, J. Mater. Review of current high-ZT thermoelectric materials. Sci. 55 (2020) 12642–12704.
[9]  M. Bala, A. Masarrat, V. Kumar, S. Ojha, K. Asokan, S. Annapoorni. Effect of thermal annealing on thermoelectric properties of Bi$_x$Sb$_{2-x}$Te$_3$ thin films grown by sputtering. J. Appl. Phys. 127 (2020), 245108.
[10] I. Malik, T. Sribastava, K.K. Surthi, C. Gayner, K.K. Kar, Enhanced thermoelectric performance of n-type Bi2Te3 alloyed with low cost and highly abundant sulfur, Mater. Chem. Phys. 255 (2020), 123598.

[11] J.L. Gao, T. Mao, T. Lv, Z.M. Li, G.Y. Xu. Thermoelectric performance of n-type $(PbTe)_{1-x}(CoTe)_x$ composite prepared by high pressure sintering method. J. Mater. Sci. -Mater. El 29 (2018) 5327–5336.

[12] N.T. Hung, E.H. Hasdeo, A.R.T. Nugraha, M.S. Dresselhaus, R. Saito. Quantum Effects in the Thermoelectric Power Factor of Low-Dimensional Semiconductors. Phys. Rev. Lett. 117 (2016), 036602.

[13] L.P. Hu, T.J. Zhu, X.H. Liu, X.B. Zhao. Point Defect Engineering of High-Performance Bismuth-Telluride-Based Thermoelectric Materials. Adv. Funct. Mater. 24 (2014) 5211–5218.

[14] B.C. Qin, Y. Xiao, Y.M. Zhou, L.D. Zhao. Thermoelectric transport properties of Pb–Sn–Te–Se system. Rare Metals 37 (2018) 343–350.

[15] S. Bathula, M. Jayasimhadri, N. Singh, A.K. Srivastava, J. Pulikkotil, A. Dhar, R. C. Budhani. Enhanced thermoelectric figure-of-merit in spark plasma sintered nanostructured n-type SiGe alloys. Appl. Phys. Lett. 101 (2012), 213902.

[16] C. Lan, A.J. Minnich, G. Chen, Z.F. Ren. Enhancement of Thermoelectric Figure-of-Merit by a Bulk Nanostructuring Approach. Adv. Funct. Mater. 20 (2010) 357–376.

[17] J.R. Sootsman, R.J. Pcionek, H.J. Kong, C. Uher, M.G. Kanatzidis. Strong Reduction of Thermal Conductivity in Nanostructured PbTe Prepared by Matrix Encapsulation. Chem. Mater. 18 (2006) 4993.

[18] A. Gueguen, P.F.P. Poudeu, C.P. Li, S. Moses, C. Uber, J.Q. He, V. Dravid, K. A. Paraskevopoulos, M.G. Kanatzidis. Thermoelectric Properties and Nanostructuring in the p-Type Materials $NaPb_{18-x}Sn_xMTe_{20}$ (M=Sb, Bi). Chem. Mater. 21 (2009) 1683–1694.

[19] G.H. Zhu, H. Lee, Y.C. Lan, X.W. Wang, G. Joshi, D.Z. Wang, J. Yang, D. Vashaee, H. Guilbert, A. Pillitteri, M.S. Dresselhaus, G. Chen, Z.F. Ren. Increased Phonon Scattering by Nanograins and Point Defects in Nanostructured Silicon with a Low Concentration of Germanium. Phys. Rev. Lett. 102 (2009), 196803.

[20] L.D. Hicks, M.S. Dresselhaus. Effect of quantum-well structures on the thermoelectric figure of merit. Phys. Rev. B 47 (1993) 12727.

[21] R. Venkatasubramanian, E. Siivola, T. Colpitts, B. O'Quinn. Thin-film thermoelectric devices with high room-temperature figures of merit. Nature 413 (2001) 597–602.

[22] T.C. Harman, M.P. Walsh, B.E. Laforge, G.W. Turner, J. Electron. Nanostructured thermoelectric materials. Mater. 34 (2005) L19–L22.

[23] A.I. Boukai, Y. Bunimovich, J. Tahir-Kheli, J.K. Yu, W.A. Goddard, J.R. Heath. Silicon nanowires as efficient thermoelectric materials. Nature 451 (2008) 168–171.

[24] B. Hinterleitner, I. Knapp, M. Poneder, Y.P. Shi, H. Muller, G. Eguchi, C. Eisenmenger-Sittner, M. Stoger-Pollach, Y. Kakefuda, N. Kawamoto. Thermoelectric performance of a metastable thin-film Heusler alloy. Nature 576 (2019) 85.

[25] 曹海山. 热电制冷技术进展与展望[J]. 制冷学报, 2022, 43 (04): 26-34.

[26] Felizco J-C, Uenuma M, Fujii M-N, Uraoka Y. Improved Thermoelectric Power Factor of $InGaZnO/SiO_2$ Thin Film Transistor via Gate-Tunable Energy Filtering. IEEE. Electron. Device. L. 2021; 42: 1236-1239.

[27] Bejenari I, Kantser V, Balandin AA. Thermoelectric properties of electrically gated bismuth telluride nanowires. Phys. Rev. B. 2010; 81: 075316.

[28] Qin D-L, Pan F, Zhou J, Xu Z-B, Deng Y. High ZT and performance controllable thermoelectric devices based on electrically gated bismuth telluride thin films. Nano Energy. 2021; 89: 106472.

[29] Yang F, Wu J, Suwardi A, Zhao Y-S, Laing B-Y, Jiang J, Xu J-W, Chi D-Z, Hippalgaonkar K, Lu J-P, Ni, Z-H. Gate-Tunable Polar Optical Phonon to Piezoelectric Scattering in Few-Layer $Bi_2O_2Se$ for High-Performance Thermoelectrics. Adv. Mater. 2021; 4: 33.

[30] Wu X-M, Gao G-Y, Hu L, Qin D. 2D Nb2SiTe4 and Nb2GeTe4: promising thermoelectric figure of merit and gate-tunable thermoelectric performance. Nanotechnology. 2021; 32: 245203.

[31] Bohang Nan, Tao Guo, Hao Deng, Guangbing Zhang, Ran Shi, Jiakai Xin, Chen Tang, Guiying Xu, Output performance improvement for thermoelectric transistor with the consideration of the Thomson effect and geometry optimization, Applied Energy, 2024, 357, 122523

[32] Nan B H, Xu G Y, Yang Q X, Zhang B, Zhou X J, Innovative design and optimized performance of thermoelectric transistor driven by the Seebeck effect, Energy Conversion and Management[J]. 2023, 283, 116880

[33] Nan Bohang, Xu Guiying, Liu Wu-Ming, Yang Quanxin, Zhang Bin, Dong Yuan, Tie Jian, Guo Tao, Zhou Xiaojing, High thermoelectric performance of PNP abrupt heterostructures by independent regulation of the electrical conductivity and Seebeck coefficient, Materials Today Communications, 2022, 31: 103343.

[34] M. Zebarjadi, A. Shakouri, K. Esfarjani. Thermoelectric transport perpendicular to thin-film heterostructures calculated using the Monte Carlo technique Phys. Rev. B 74(2006), 195331.

[35] V.N. Agarev. Nonstationary thermoelectric power in multilayered structures with p-n junctions. Semiconductors 31 (1997) 784.

[36] K. Kim, J. Park, S.U. Kim, O. Kwon, J.S. Lee, S.H. Park, Y.K. Choi. Thermopower profiling of a silicon p-n Junction. Appl. Phys. Lett. 90 (2007) 43107.

[37] X.Y. Wang, H.J. Wang, B. Xiang, L.W. Fu, H. Zhu, D. Chai, B. Zhu, Y. Yu, N. Gao, Z. Y. Huang, F.Q. Zu. Thermoelectric Performance of $Sb_2Te_3$-Based Alloys is Improved by Introducing PN Junctions. ACS Appl. Mater. Interfaces 10 (2018) 23277.

[38] I.I. Balmush, Z.M. Dashevskii, A.I. Kasiyan. Potential Distribution in Semiconductor Structure with p−n Junction in the Presence of Temperature Gradient. Fiz. i Tekh. Poluprovodn. 20 (1986)1541 (In Russian).

[39] Z.M. Dashebsky, S. Ashmontas, L. Vingelis, I. Gradauskas, AI. Kasian. The thermoelectric power on p-n junction. In 15th Int. Conf. on Thermoelectrics. IEEE. 336 (1996).

[40] A.A. Zakhidov, Y.I. Ravich, D.A. Pchenoy-Severin. Thermopower enhancement and optimal ZT in p-n junction arrays. In 18th Int. Conf. on Thermoelectrics. IEEE. 193 (1999).

[41] Y.I. Ravich, D.A. Pshenai-Severin. Thermoelectric figure of merit of a p-n junction. Semiconductors 35 (2001) 1161–1165.

[42] Chen Z W, Jian Z Z, Li W, et al. Lattice dislocations enhancing thermoelectric PbTe in addition to band convergence [J]. Advanced Materials, 2017, 29(23): 1606768.

[43] He J, Xu J T, Liu G Q, et al. Enhanced thermopower in rock-salt SnTe–CdTe from band convergence [J]. RSC Advances, 2016, 6(38): 32189-32192.

[44] V. A. Kulbachinskii, M. Inoue, M. Sasaki, H. Negishi, W. X. Gao, K. Takase, Y. Giman, P. Lostak, and J. Horak. Valence-band changes in $Sb_{2-x}In_xTe_3$ and $Sb_2Te_{3-y}Se_y$ by transport and Shubnikov–de Haas effect measurements. Phys. Rev. B 50, 16921

[45] Xu G-Y, Ren P, Lin T, Wu X-F, Zhang Y-H, Niu S-T, Bailey TP. Mechanism and application method to analyze the carrier scattering factor by electrical conductivity ratio based on thermoelectric property measurement. J. Appl. Phys. 2018; 123: 015101.

[46] Wang S-Y, Tan G-J, Xie W-J, Zheng G, Li H, Yang J-H, Tang X-F. Enhanced thermoelectric properties of $Bi_2(Te_{1-x}Se_x)_3$-based compounds as n-type legs for low temperature power generation. J. Mater. Chem. 2012; 22: 20943.

[47] Dai W, Liu W-K, Yang J, Xu C, Alabastri A, Liu C, Nordlander P, Guan Z-Q, Xu H-X. Giant photothermoelectric effect in silicon nanoribbon photodetectors. Light Sci. Appl. 2020; 9: 120.

[48] Ge Y, He K, Xiao L-H, Yuan W-Z, Huang S-M. Geometric optimization for the thermoelectric generator with variable cross-section legs by coupling finite element method and optimization algorithm. Renew Energ. 2022; 183: 294.

[49] Fu D, Levander A-X, Zhang R, Ager III J-W, Wu J. Electrothermally driven current vortices in inhomogeneous bipolar semiconductors. Phys. Rev. B. 2011; 84: 045205.

[50] Lv T, Li Z-M, Yang Q-X, Benton A, Zheng H-T, Xu G-Y. Synergistic regulation of electrical-thermal effect leading to an optimized thermoelectric performance in Co doping n-type $Bi_2(Te_{0.97}Se_{0.03})_3$. Intermetallics. 2020; 118: 106683.

[51] Chavez R, Angst S, Hall J, Stoetzel J, Kessler V, Bitzer L, Maculewicz F, Benson N, Wiggers H, Wolf D, Schierning G, Schmechel R. High Temperature Thermoelectric Device Concept Using Large Area PN Junctions. J. Electron. Mater. 2014; 43: 2376.